\documentclass[preprint2,numberedappendix]{emulateapj-rtx4}
\usepackage{graphicx,times,bm,url,color}
\graphicspath{{./fig/}{./png/}}

\newcommand{\brac}[1]{\langle #1 \rangle}
\newcommand{\EQ}{\begin{equation}}
\newcommand{\EN}{\end{equation}}
\newcommand{\EQA}{\begin{eqnarray}}
\newcommand{\ENA}{\end{eqnarray}}

\newcommand{\Eq}[1]{Equation~(\ref{#1})}

\newcommand{\Fig}[1]{Figure~\ref{#1}}

\newcommand{\mean}[1]{\overline #1}

{}
{}
{}

{}
{}
{}
{}
{}
{}
{}
{}
{}
{}
{}
{}
{}
{}
{}
{}
{}

{}
{}
{}

\newcommand{\meanR}{\overline{\rho}}

{}

{}
{}

\newcommand{\eee}{\hat{\mbox{\boldmath $e$}} {}}

\newcommand{\uu}{\mbox{\boldmath $u$} {}}

\newcommand{\ssss}{\mbox{\boldmath $\xi$} {}}
\newcommand{\bb}{\mbox{\boldmath $b$} {}}

\newcommand{\jj}{\mbox{\boldmath $j$} {}}

\newcommand{\nab}{\mbox{\boldmath $\nabla$} {}}
\newcommand{\nablar}{\nabla_{\!r}}

\newcommand{\oo}{\mbox{\boldmath $\omega$} {}}

\def\Pra{\mbox{\rm Pr}}

\def\Pm{\mbox{\rm Pr}_M}

\def\urms{u_{\rm rms}}

\def\etatz{\eta_{\rm t0}}

\def\Btt{\mean{B_\phi}^{\rm rms}}
\def\Brt{\mean{B_r}^{\rm rms}}

\newcommand{\chitm}{\overline{\chi}_{\rm SGS}}
\newcommand{\Prat}{\Pra_{\rm SGS}}
\newcommand{\tauc}{\tau_{\rm c}}

\newcommand{\s}{\,{\rm s}}
\begin{document}

\title{ON THE CAUSE OF SOLAR-LIKE EQUATORWARD MIGRATION IN GLOBAL\\ CONVECTIVE
  DYNAMO SIMULATIONS}

\author{J\"orn Warnecke$^{1,2}$, Petri J.\ K\"apyl\"a$^{2,3}$, Maarit
J.\  K\"apyl\"a$^{2}$, and Axel Brandenburg$^{4,5}$}
\affil{
$^1$Max-Planck-Institut f\"ur Sonnensystemforschung,
  Justus-von-Liebig-Weg 3, D-37077 G\"ottingen, Germany; warnecke@mps.mpg.de\\
$^2$ReSoLVE Centre of Excellence, Department of Information and
Computer Science, Aalto University, P.O. Box 15400, FI-00\ 076 Aalto,
Finland\\
$^3$Physics Department, Gustaf H\"allstr\"omin katu 2a, P.O. Box 64,
FI-00014 University of Helsinki, Finland\\
$^4$NORDITA, KTH Royal Institute of Technology and Stockholm University,
Roslagstullsbacken 23, SE-10691 Stockholm, Sweden\\
$^5$Department of Astronomy, AlbaNova University Center,
Stockholm University, SE-10691 Stockholm, Sweden
}
\email{warnecke@mps.mpg.de
($ $Revision: 1.181 $ $)
}

\begin{abstract}
We present results from four convectively-driven stellar dynamo
simulations in spherical wedge geometry.
All of these simulations produce cyclic and migrating mean magnetic fields.
Through detailed comparisons, we show that
the migration direction can be explained by an
$\alpha\Omega$ dynamo wave following the Parker--Yoshimura rule.
We conclude that the equatorward migration in this and previous work
is due to a positive (negative) $\alpha$ effect in the northern
(southern) hemisphere and a negative radial gradient of $\Omega$
outside the inner tangent cylinder of these models.
This idea is supported by a strong correlation between negative radial 
shear and toroidal field strength in the region of equatorward propagation.
\end{abstract}

\section{Introduction}
\label{intro}

Just over 50 yr after the paper by \cite{Mau04}, in which he showed
for the first time the equatorward migration (EM) of sunspot activity
in a time--latitude (or butterfly) diagram, \cite{P55} proposed
a possible solution: migration of an $\alpha\Omega$ dynamo wave along
lines of constant
angular velocity $\Omega$ \citep{Yos75}.
Here, $\alpha$ is related to kinetic helicity and is positive (negative)
in the northern (southern) hemisphere \citep{SKR66}.
To explain EM, $\nab\Omega$ must point in the negative radial direction.
However, application to the Sun became problematic with the advent of
helioseismology showing that $\nablar\Omega$ is actually positive at
low latitudes where sunspots occur \citep{Schouea98}, implying poleward
migration (PM).
This ignores the near-surface shear layer where a negative
$\nablar\Omega$ \citep{Thompson96} could cause EM \citep{B05}.
An alternative solution was offered by \cite{CSD95}, who found that in
$\alpha\Omega$ dynamo models with spatially separated induction layers
the direction of migration can also be controlled by the
direction of meridional circulation at the bottom of the convection
zone, where the observed poleward flow at the surface must lead to an
equatorward return flow.
Finally, even with just uniform rotation, i.e., in an $\alpha^2$ dynamo
as opposed to the aforementioned $\alpha\Omega$ dynamos, it may be
possible to obtain EM due to the fact that $\alpha$ changes sign at the
equator \citep{BS87,Ra87,MTKB10,WBM11}.

Meanwhile, global dynamo simulations driven by rotating convection in
spherical shells have demonstrated not only the production of large-scale
magnetic fields, but, in some cases, also EM
\citep{KMB12,KMCWB13,WKMB13b,ABMT13}.
Although this seemed to be successful in reproducing Maunder's
observation of EM,
the reason remained unclear.
Noting the agreement between their simulation and the $\alpha^2$ dynamo of
\cite{MTKB10} in terms of the $\pi/2$ phase shift between poloidal and
toroidal fields near the surface, as well as their similar amplitudes, \cite{KMCWB13}
suggested such an $\alpha^2$ dynamo as a possible underlying mechanism.
Yet another possibility is that $\alpha$ can change sign
if the second term in the estimate for $\alpha$ \citep{PFL76},
\EQ
\alpha={\tauc\over3}\left(-\overline{\oo\cdot\uu}
+{\overline{\jj\cdot\bb}\over\meanR}\right),
\label{alpha}
\EN
becomes dominant near the surface, where the mean density $\meanR$
becomes small.
Here, $\oo=\nab\times\uu$ is the vorticity, $\uu$ is the small-scale
velocity,
$\jj=\nab\times\bb/\mu_0$ is the current density, $\bb$ is the small-scale magnetic field,
$\mu_0$ is the vacuum permeability, 
$\tauc$ is the correlation time of the turbulence,
and overbars denote suitable (e.g., longitudinal) averaging.
However, an earlier examination by \cite{WKMB13a} showed that the
data do not support this idea, i.e., the contribution from the second term
is not large enough.
Furthermore, the theoretical justification for \Eq{alpha} is questionable \citep{BRRS08}.

A potentially important difference between the models of \cite{KMB12,KMCWB13}
and those of other groups
\citep{GCS10,Racine11,BMBBT11,ABBMT12,ABMT13,NBBMT13}
is the use of a blackbody condition for the entropy and a
radial magnetic field on the outer boundary.
The latter may be more realistic for the solar surface
\citep{CSJI12}.

It should be noted that a near-surface negative shear layer similar to
the Sun was either not resolved in the simulations of
\cite{KMB12,KMCWB13}, or,
in the case of \cite{WKMB13b}, such a layer did not coincide with the
location of EM.
Instead, most of these simulations show a strong tendency for the contours of
angular velocity to be constant on cylinders.
Some of them even show a local minimum of angular velocity at mid-latitudes.
Indeed, \cite{ABMT13} identified EM with the location of the greatest
latitudinal shear at a given point in the cycle and find that also weak
negative radial shear plays a role.
latitudinal shear at a given point in the cycle and find that also weak
negative radial shear plays a role.

In this Letter, we show through detailed comparison
among four models that it is this local minimum, where
$\nablar\Omega<0$ and $\alpha >0$, which explains the EM as a Parker
dynamo wave traveling equatorward.
While we do not expect this to apply to Maunder's observed EM in the Sun,
it does clarify the outstanding question regarding the origin of EM
in the simulations.
A clear understanding of these numerical experiments is a prerequisite
for a better understanding of the processes causing EM in the Sun.

\section{Strategy}

\begin{figure}[t!]
\begin{center}
\includegraphics[width=0.4\textwidth]{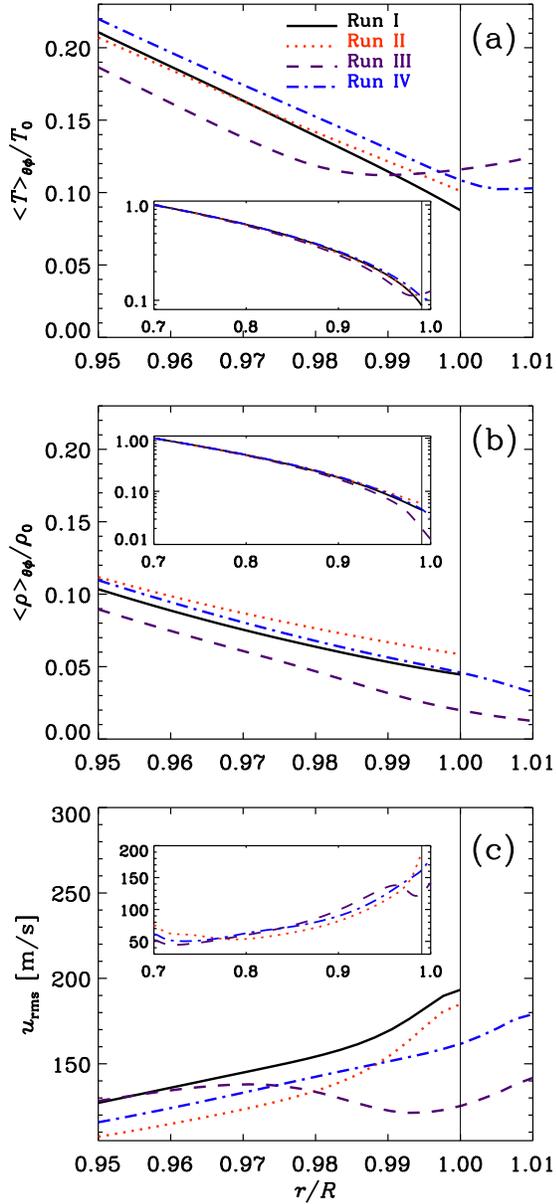}
\end{center}\caption[]{
Radial profiles of azimuthally and latitudinally averaged temperature
$\brac{T}_{\theta\phi}$ (a), density $\brac{\rho}_{\theta\phi}$ (b), and rms velocity $\urms$ (c)
near the surface, normalized by their values at the bottom of
the domain $T_0$, $\rho_0$, or in m/s respectively.
The inlays show the entire radial extent.
The solid black lines indicate Run~I, red dotted Run~II, purple dashed
Run~III, and blue dash-dotted Run~IV.
The thin black lines represent the surface ($r=R$).
}
\label{strat}
\end{figure}

To isolate effects arising from changes in the $\alpha$ effect and the
differential rotation, we consider four models.
Our reference model, {\bf Run~I}, is the same model presented in
\cite{KMB12} as their Run B4m and in as \cite{KMCWB13} their Run C1.
{\bf Run~II} is a run in which the subgrid scale (SGS) Prandtl number,
$\Prat=\nu/\chitm$, is reduced
from 2.5 to 0.5, and the magnetic Prandtl number, $\Pm=\nu/\eta$, is
reduced from 1 to 0.5.
Here $\nu$ is the viscosity, $\eta$ is the
magnetic diffusivity, and $\chitm$ is the mean SGS heat diffusivity.
We keep $\nu$ fixed so the effect of lowering $\Prat$ is that the SGS 
diffusion is more efficiently smoothing out entropy variations.
For stars, the relevant value of $\Prat$ is well below unity, but 
such cases are difficult to simulate numerically.
In {\bf Runs~III} and {\bf IV}, we have replaced the
outer radiative boundary condition by a cooling layer \citep[see][for
the implementation and the profile]{WKMB13b} above fractional radii
$r/R=0.985$ and $1.0$, respectively; see
\Fig{strat}(a) and (b) for the radial temperature and density profiles.
Here, $R$ is the solar radius.
The cooling profile of Run~III leads to a stronger density
decrease and suppression of $\urms$ than in the other runs; see
\Fig{strat}(b) and (c).
Besides the differences in the fluid and magnetic Prandtl numbers (Run~II)
and in the upper thermal boundary condition (Runs~III and IV), the setups are equal.
We can therefore isolate the origin of the difference in the
migration pattern of the toroidal field.

Our simulations are done in a wedge $|90^\circ-\theta|\leq 75^\circ$,
$0<\phi<90^\circ$, and $R-\Delta R\leq r\leq R+\delta R_C$,
where $\theta$ is colatitude, $\phi$ is longitude, $r$ is radius,
$\Delta R=0.3\,R$ and $\delta R_C=0.01\,R$ is the extension by the
cooling layer in Runs~III and IV.
We solve the equations of compressible magnetohydrodynamics
using the {\sc Pencil Code}.\footnote{http://pencil-code.google.com/}
The basic setup of these four models is identical to previous work;
the details can be found in \cite{KMCWB13} and \cite{WKMB13b}.
We scale our results to physical units following
\cite{KMCWB13,KMB14} and choose a rotation rate of
$\Omega_0=5\Omega_\sun$, where $\Omega_\sun=2.7\times10^{-6} \s$
is the solar value.

\section{Results}

\begin{figure*}[t!]
\begin{center}
\includegraphics[width=0.38\textwidth]{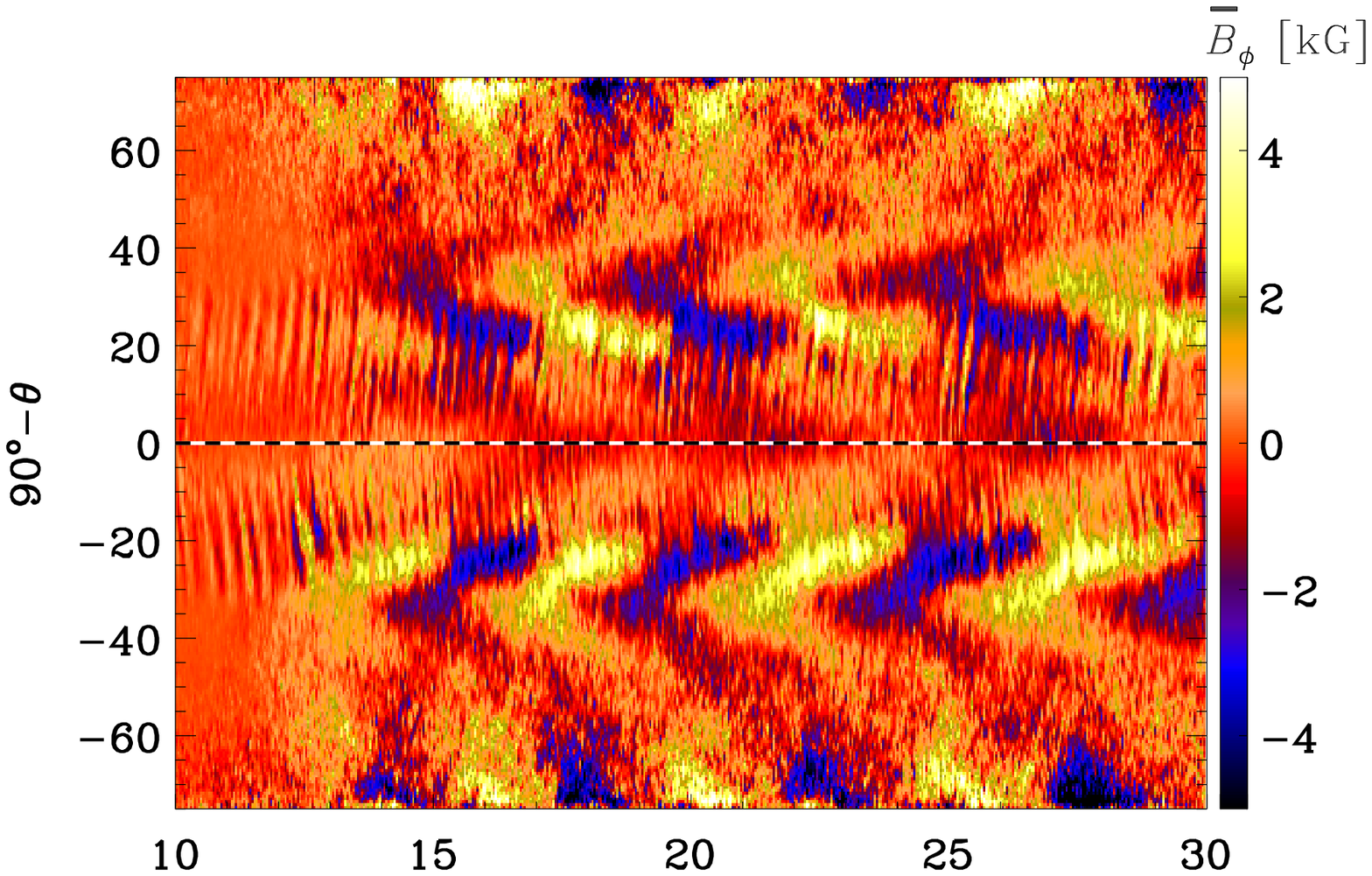}
\includegraphics[width=0.38\textwidth]{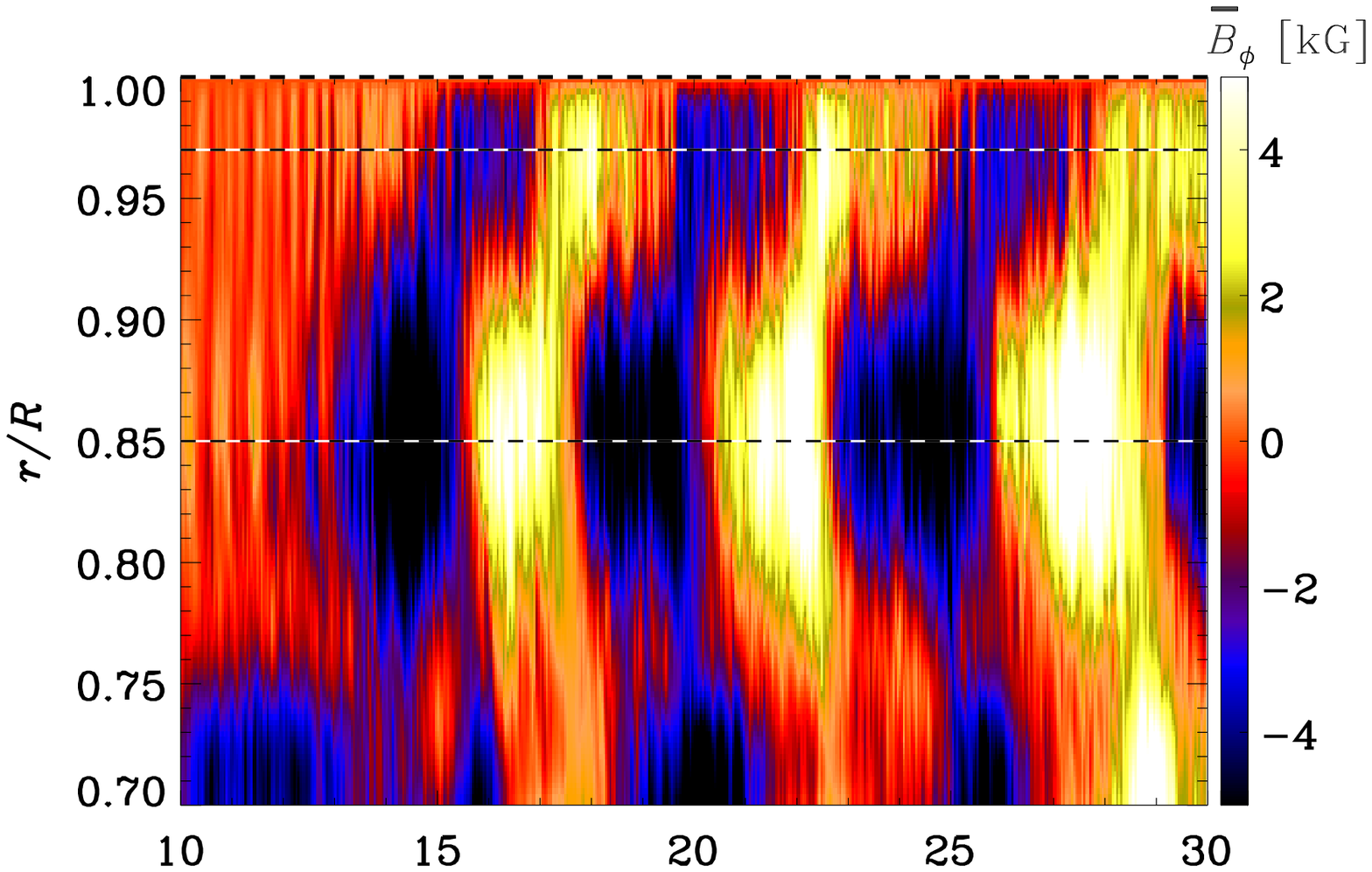}
\includegraphics[width=0.38\textwidth]{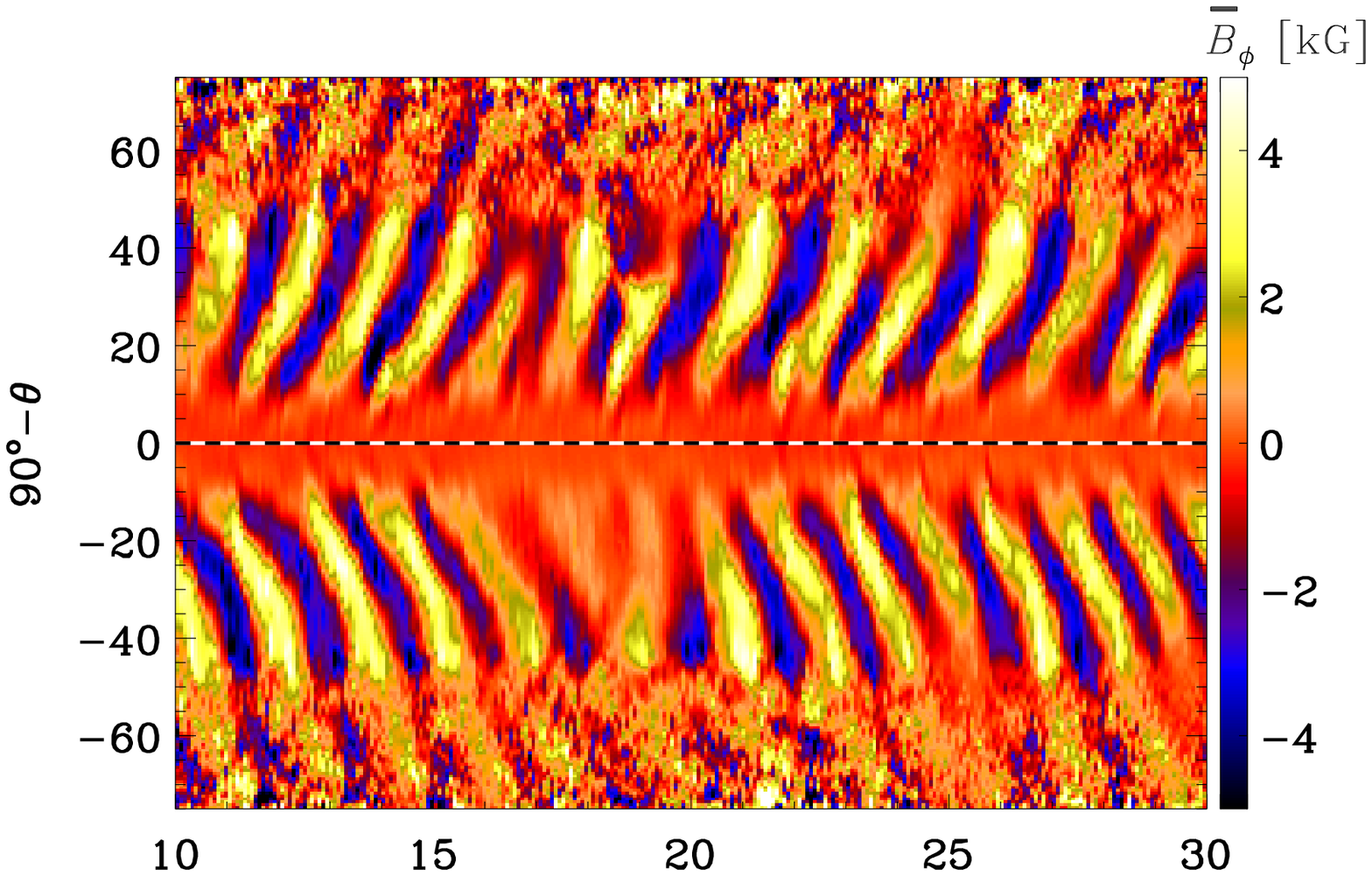}
\includegraphics[width=0.38\textwidth]{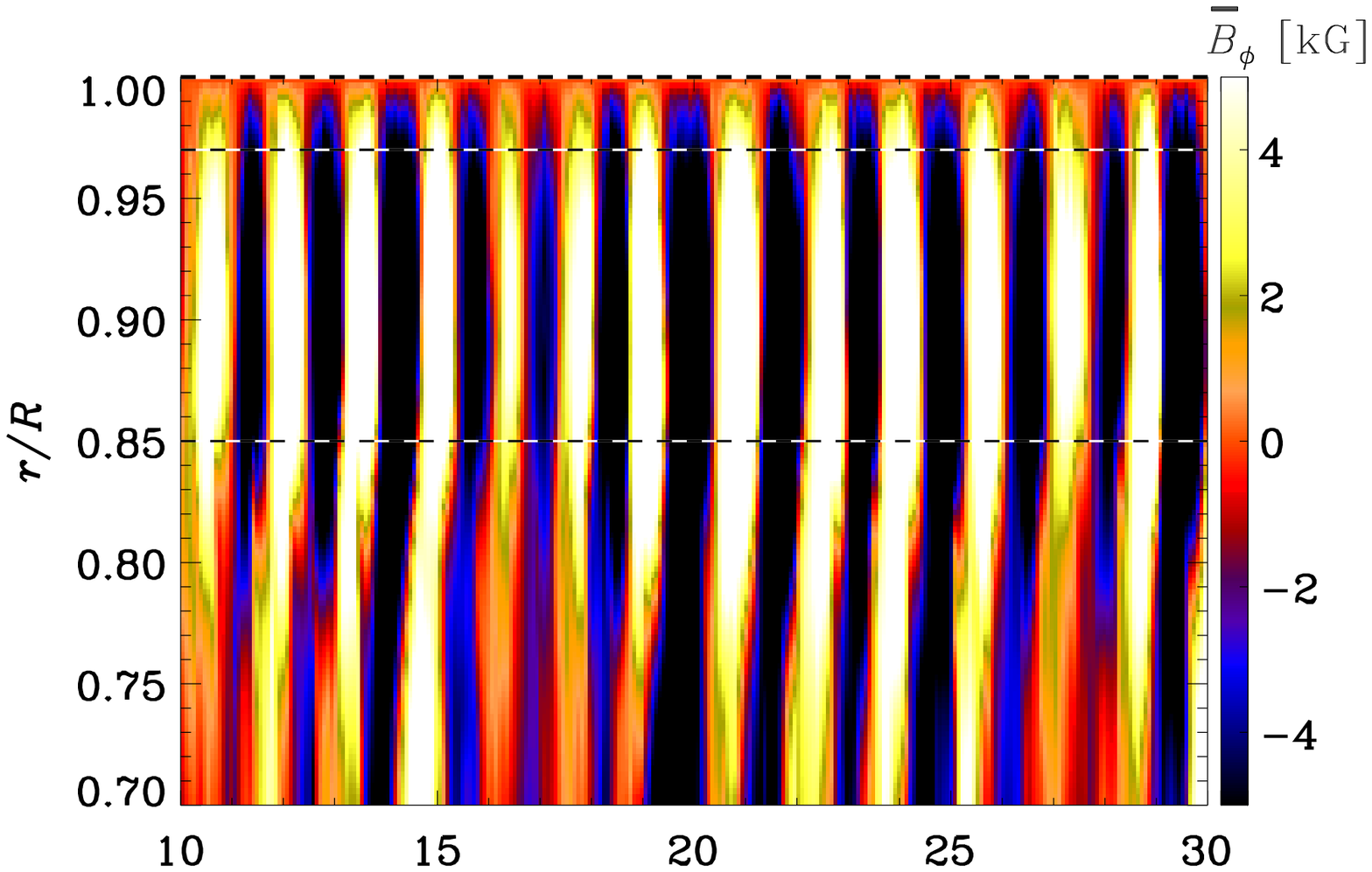}
\includegraphics[width=0.38\textwidth]{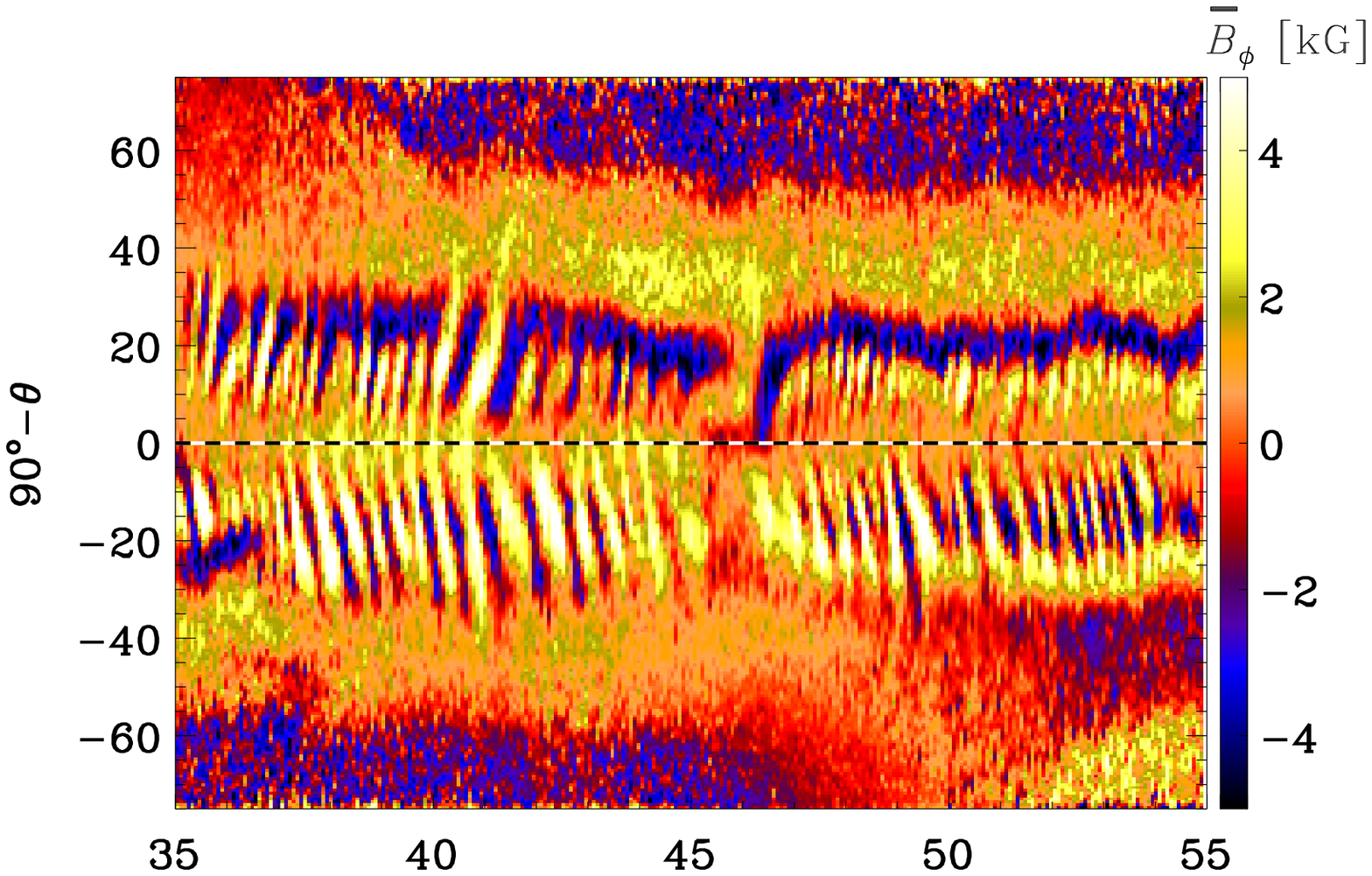}
\includegraphics[width=0.38\textwidth]{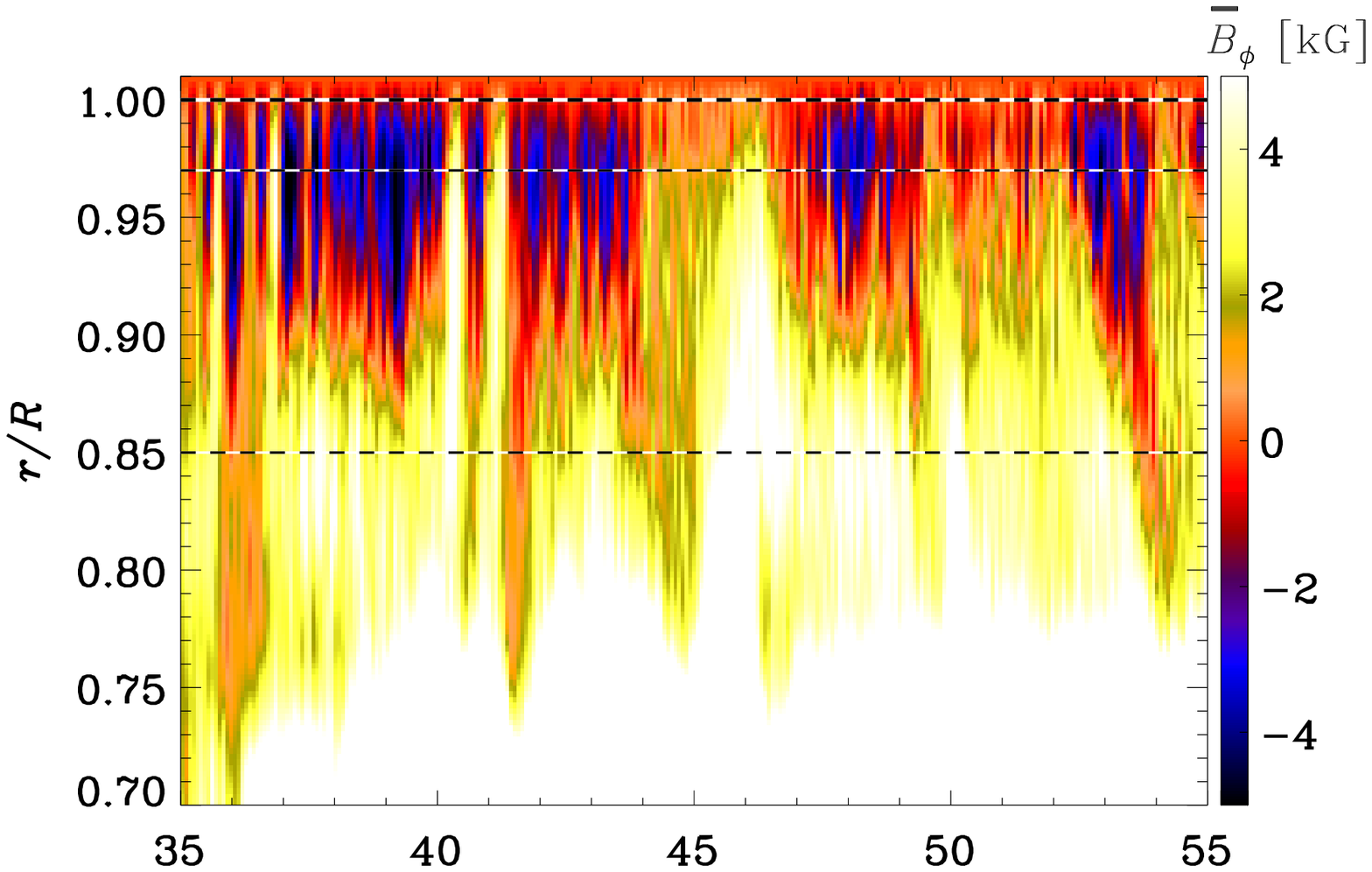}
\includegraphics[width=0.38\textwidth]{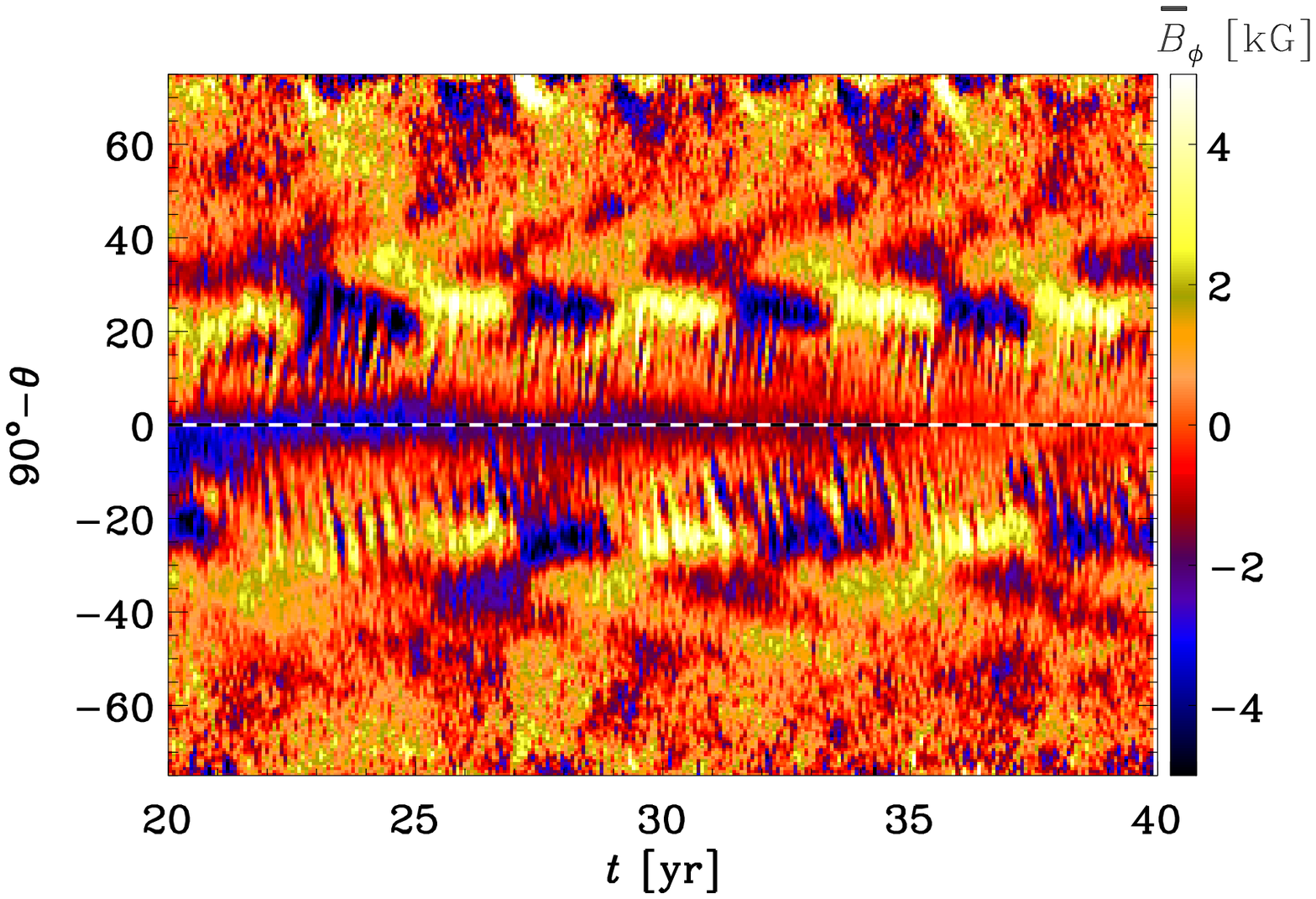}
\includegraphics[width=0.38\textwidth]{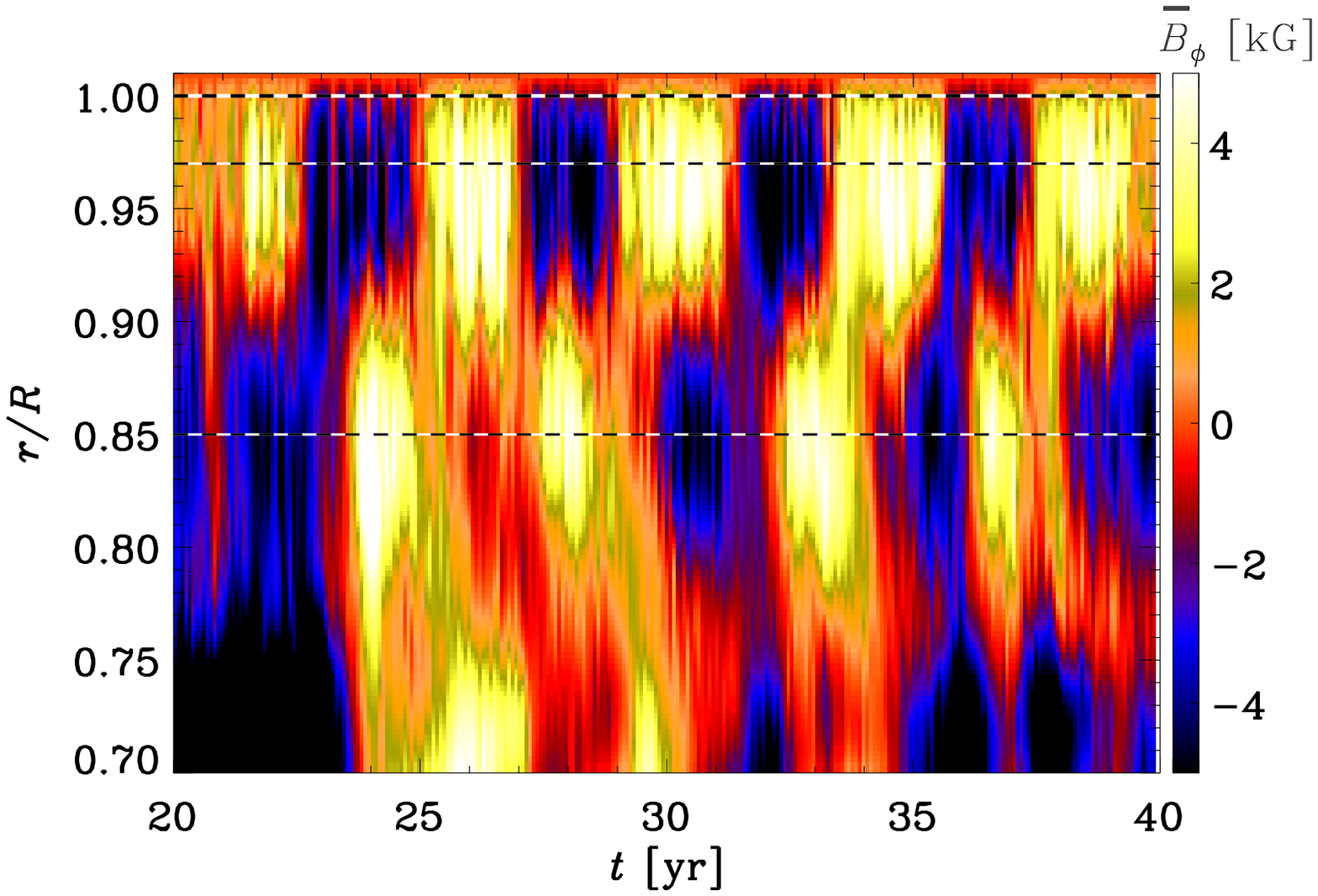}
\end{center}\caption[]{
Time evolution of the mean toroidal magnetic field $\mean{B_\phi}$ in
the convection zone for Runs~I, II, III, and IV, from top to bottom.
In the left column, the radial cut is shown at $r=0.98\,R$, and, in
the right column, the latitudinal cut at $90-\theta=25^{\circ}$.
The dashed horizontal lines show the location of the equator at
$\theta=\pi/2$ (left) and the radii $r=R$, $r=0.98\,R$ and $r=0.85\,R$ (right).
}
\label{but}
\end{figure*}

\begin{figure*}[t!]
\begin{center}
\includegraphics[width=0.247\textwidth]{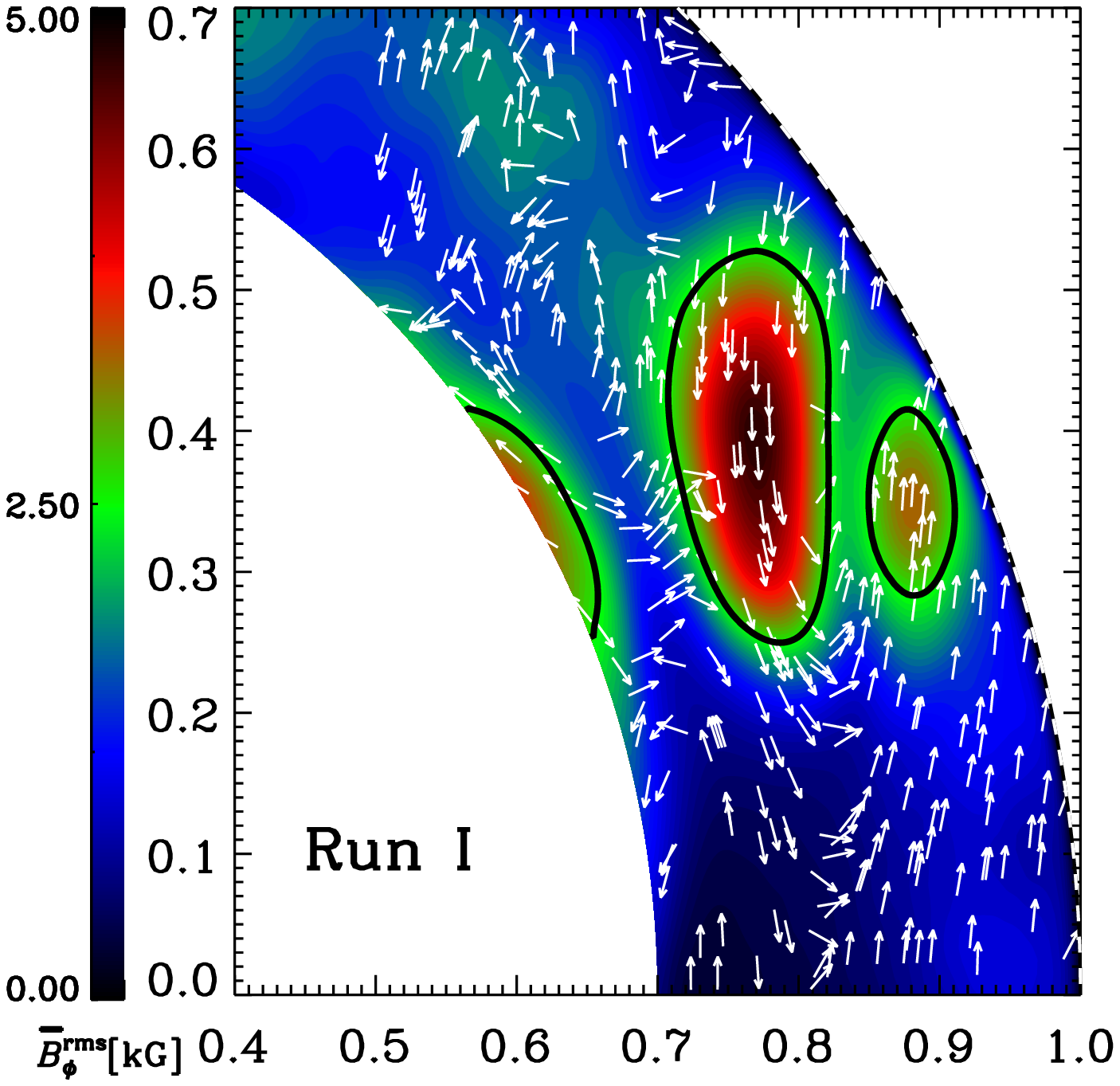}
\includegraphics[width=0.247\textwidth]{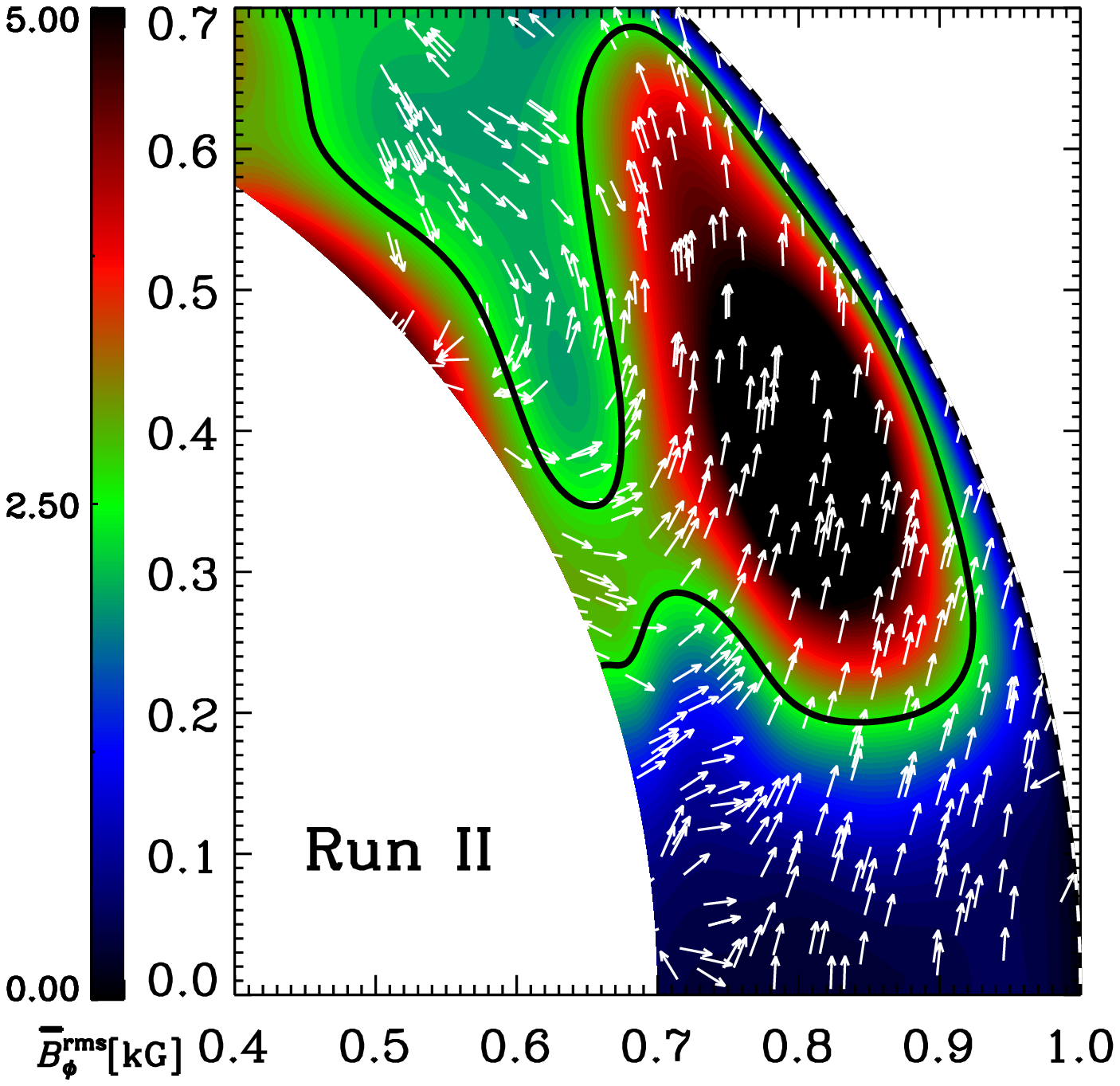}
\includegraphics[width=0.247\textwidth]{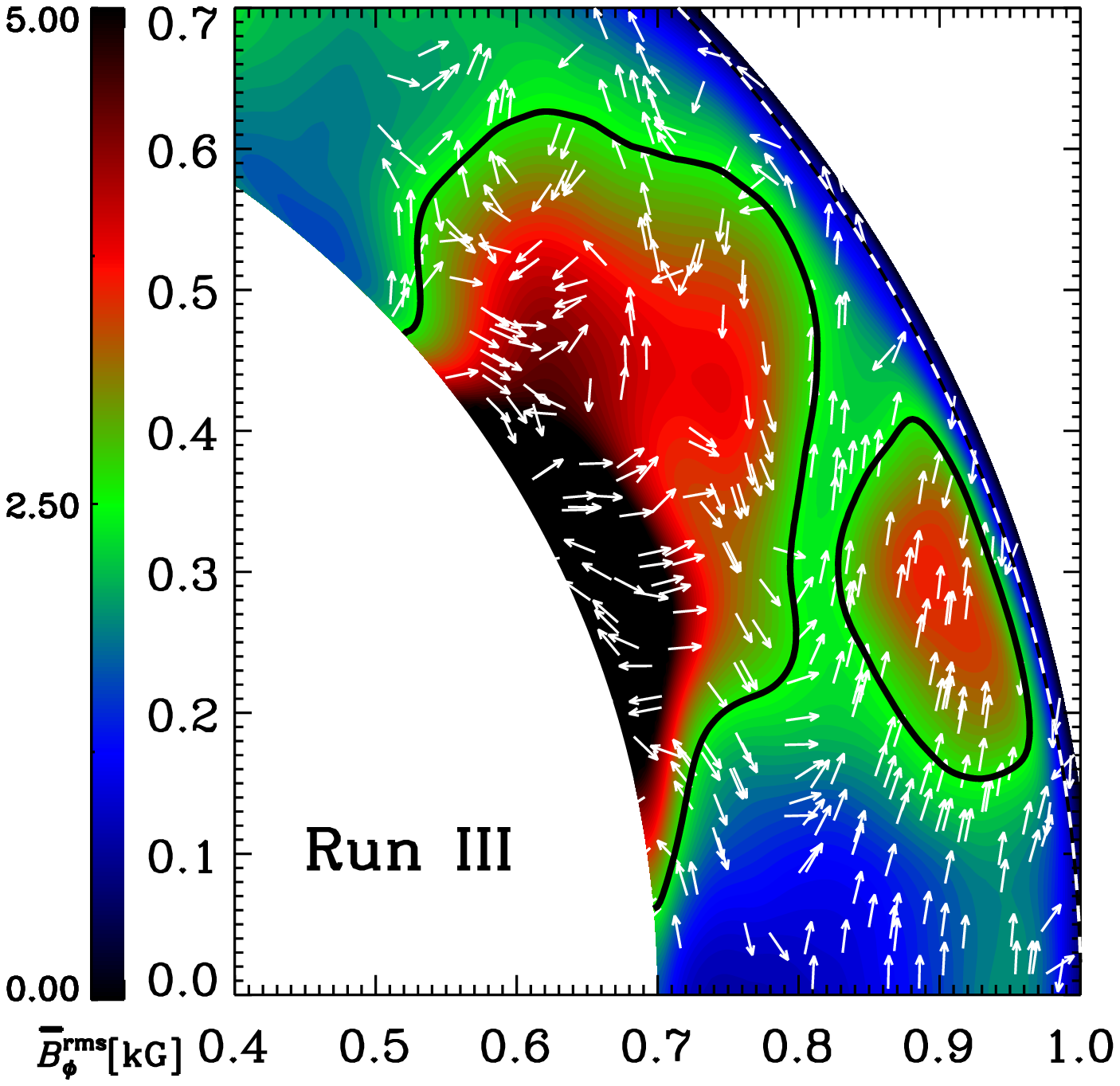}
\includegraphics[width=0.247\textwidth]{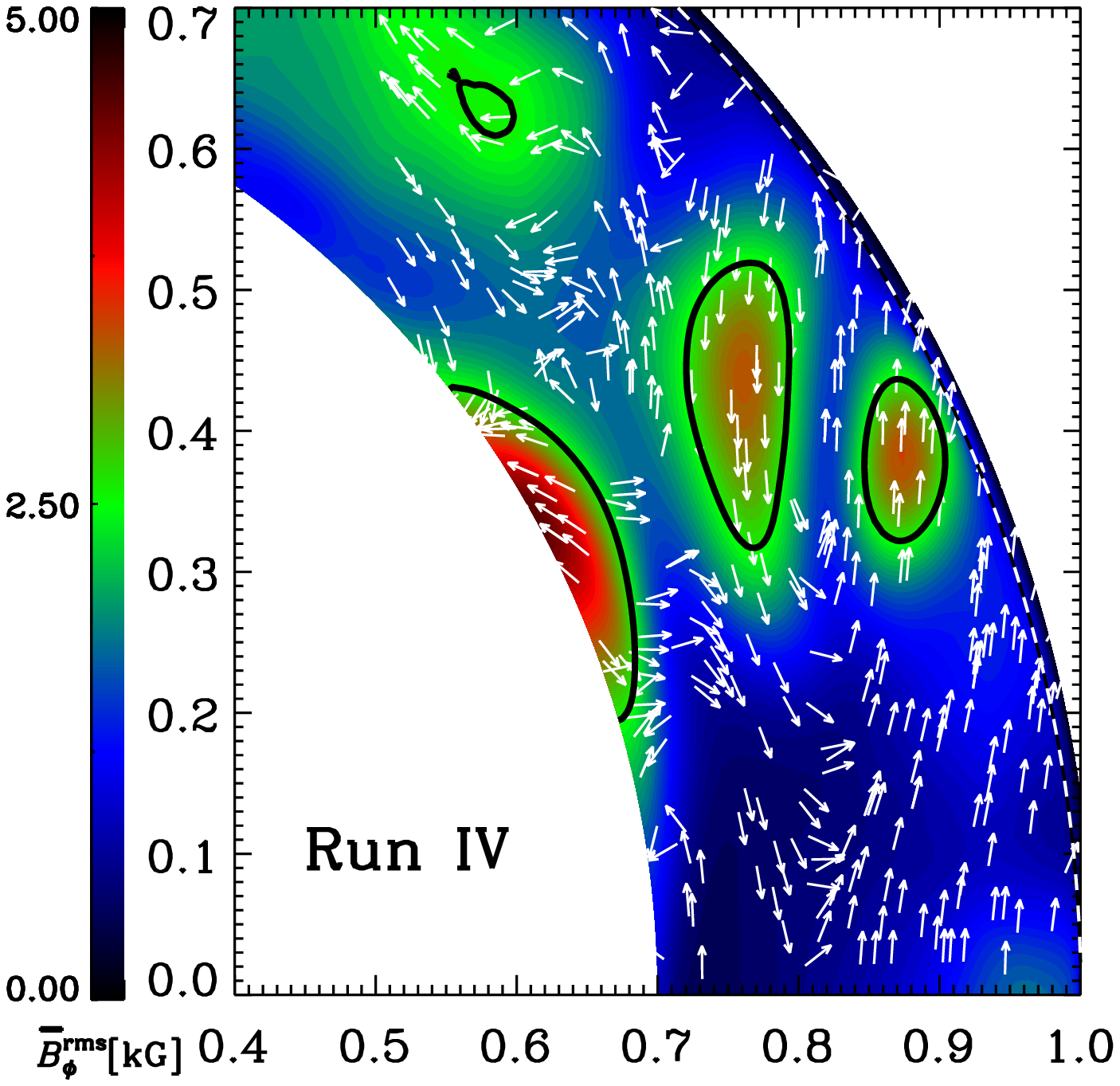}
\includegraphics[width=0.247\textwidth]{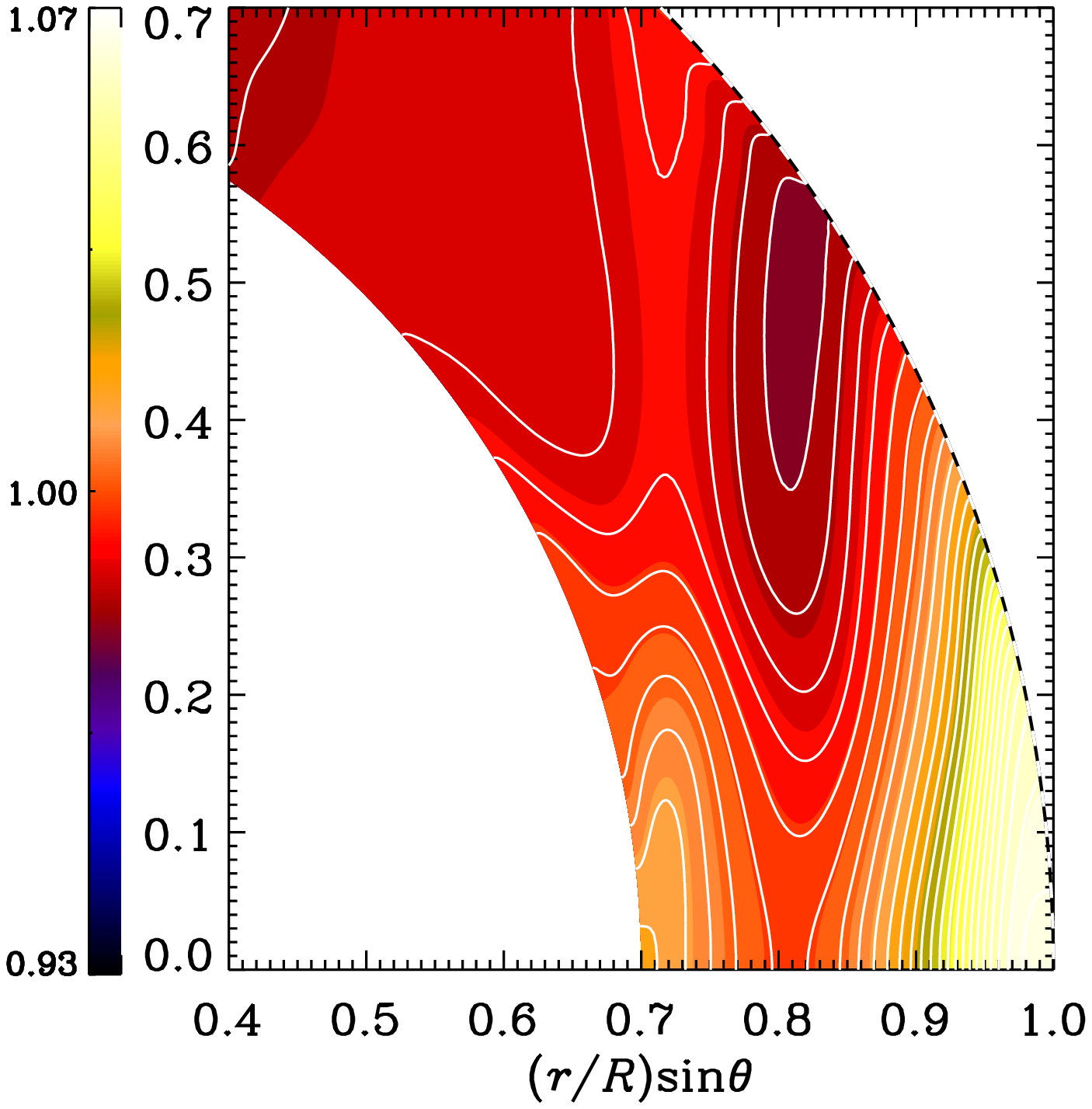}
\includegraphics[width=0.247\textwidth]{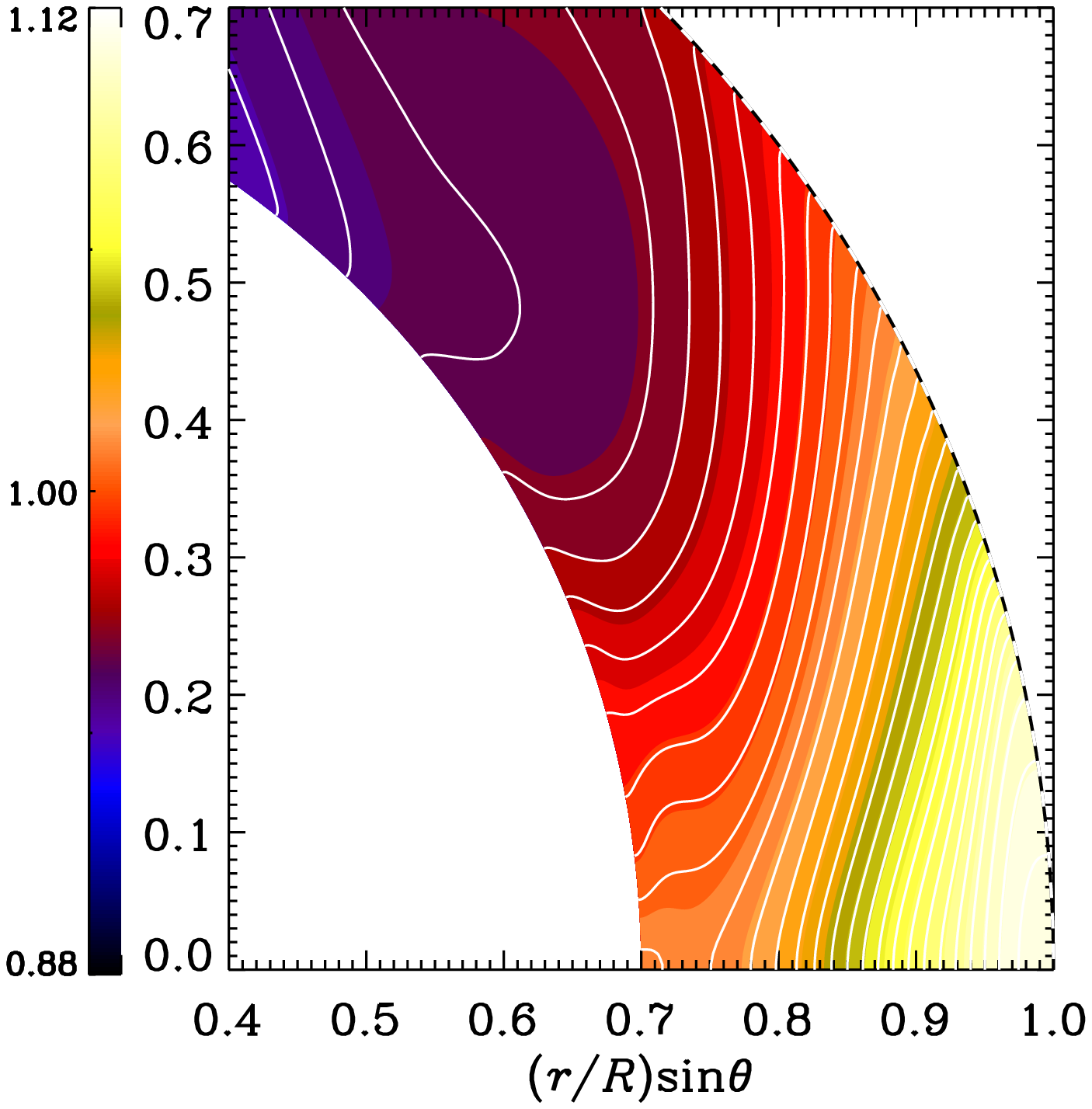}
\includegraphics[width=0.247\textwidth]{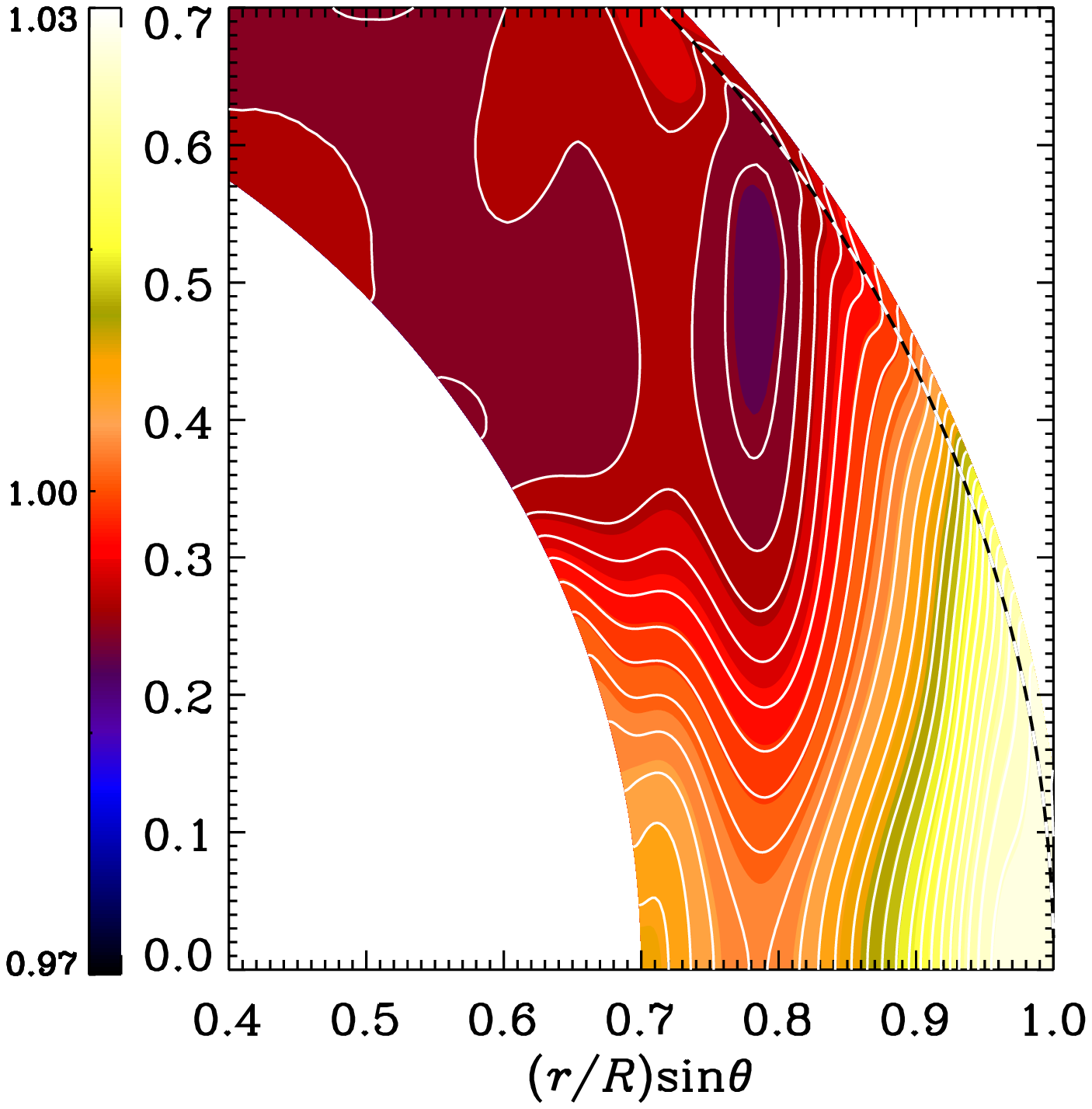}
\includegraphics[width=0.247\textwidth]{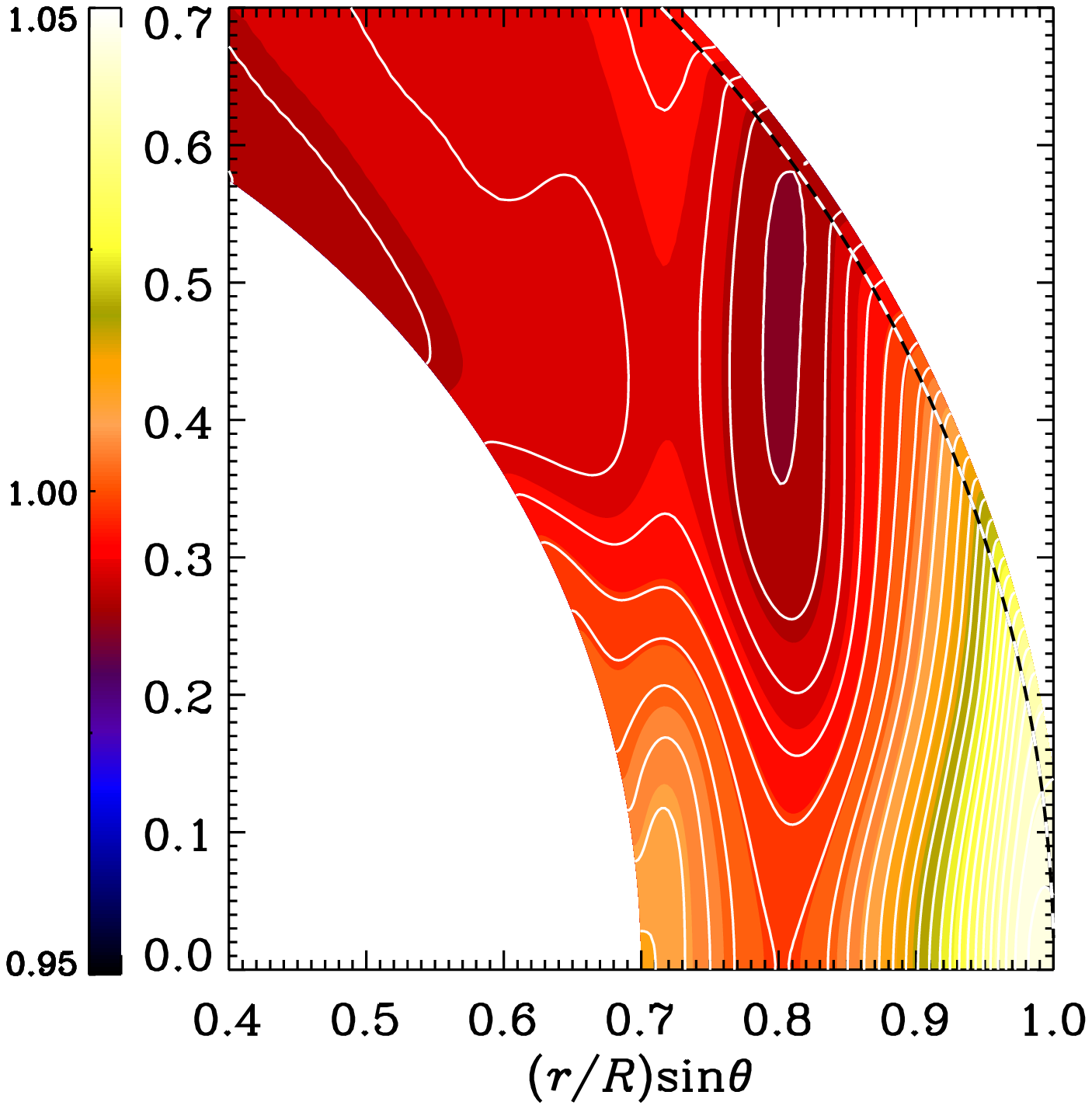}
%\includegraphics[width=0.24\textwidth]{eqmigr_1.eps}
%\includegraphics[width=0.24\textwidth]{eqmigr_1P.eps}
%\includegraphics[width=0.24\textwidth]{eqmigr_1C.eps}
%\includegraphics[width=0.24\textwidth]{eqmigr_1C2.eps}
%\includegraphics[width=0.24\textwidth]{diff_conv_1.eps}
%\includegraphics[width=0.24\textwidth]{diff_conv_1P.eps}
%\includegraphics[width=0.24\textwidth]{diff_conv_1C.eps}
%\includegraphics[width=0.24\textwidth]{diff_conv_1C2.eps}
%APJ
\end{center}\caption[]{
{\it Top row:}
color coded $\Btt$ during the saturated stage for Runs~I--IV (left to right).
White arrows show the direction of migration
$\ssss_{\rm mig}(r,\theta)=-\alpha \eee_\phi \times\nab\Omega$; see
\Eq{eqm}.
The black solid lines indicate isocontours of $\mean{B_\phi}$ at 2.5\,kG.
{\it Bottom row:}
$\Omega(r,\theta)/\Omega_0$ for the same runs.
The dashed lines indicate the surface ($r=R$).
}\label{Diffrot}
\end{figure*}

\begin{figure*}[t!]
\begin{center}
\includegraphics[width=0.4\textwidth]{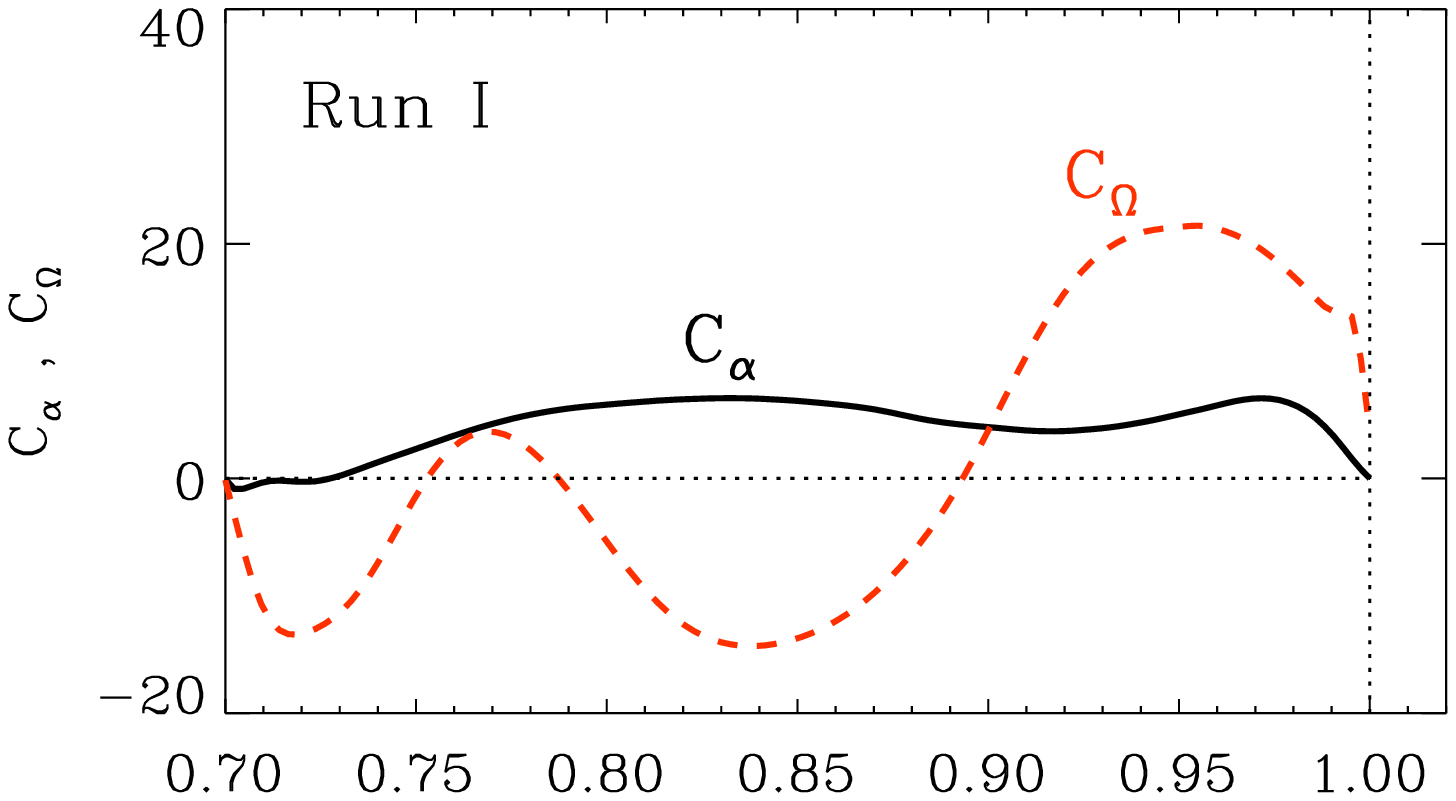}
\includegraphics[width=0.4\textwidth]{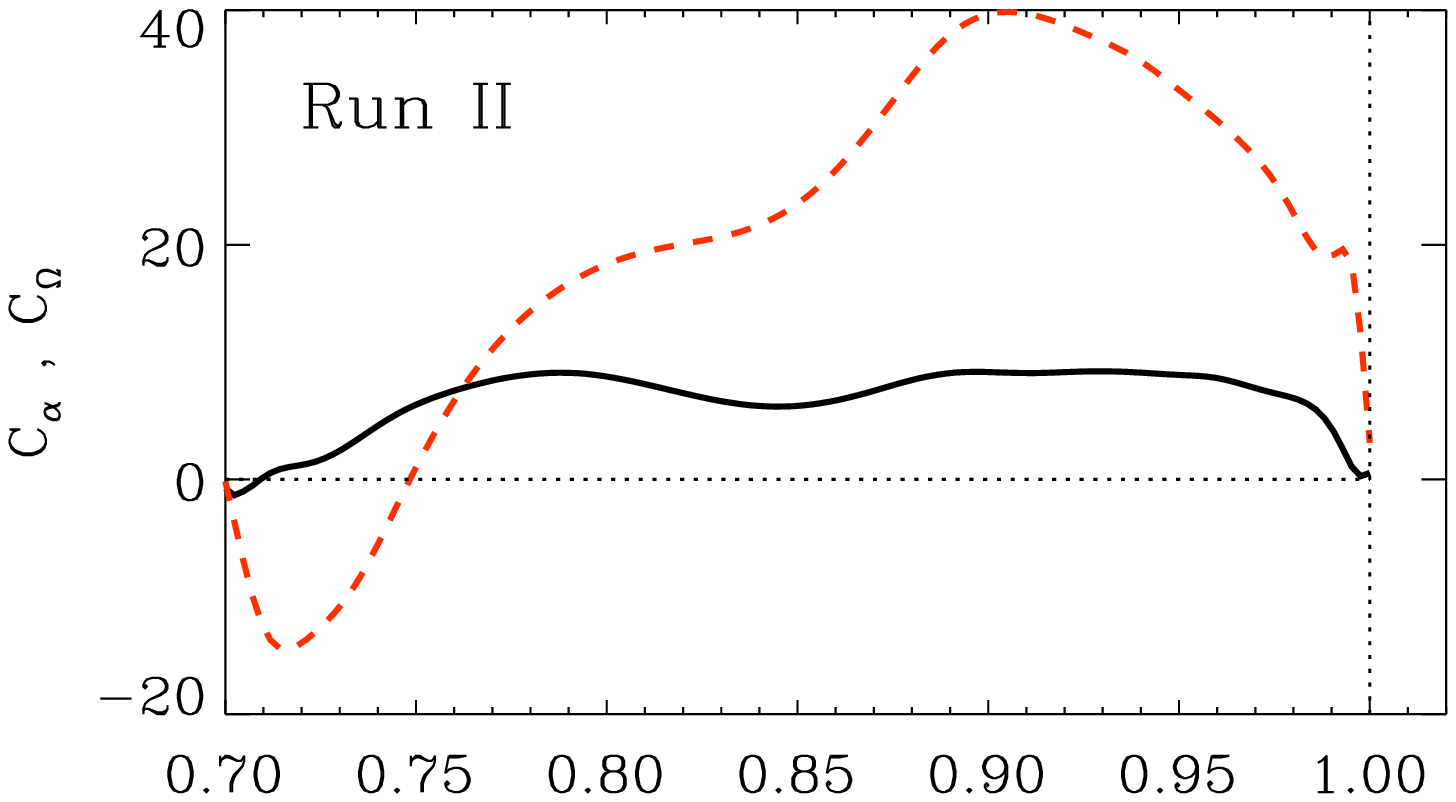}
\includegraphics[width=0.4\textwidth]{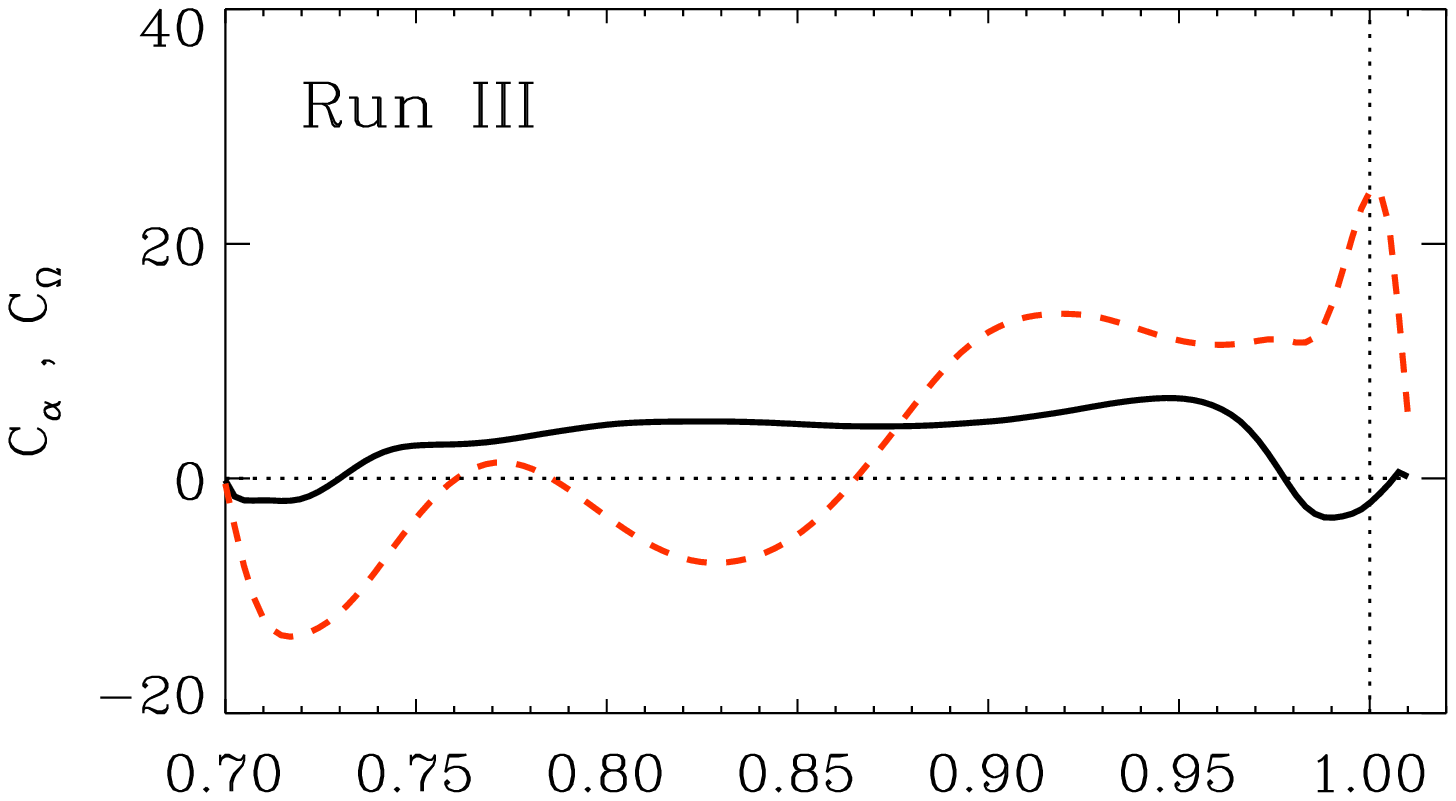}
\includegraphics[width=0.4\textwidth]{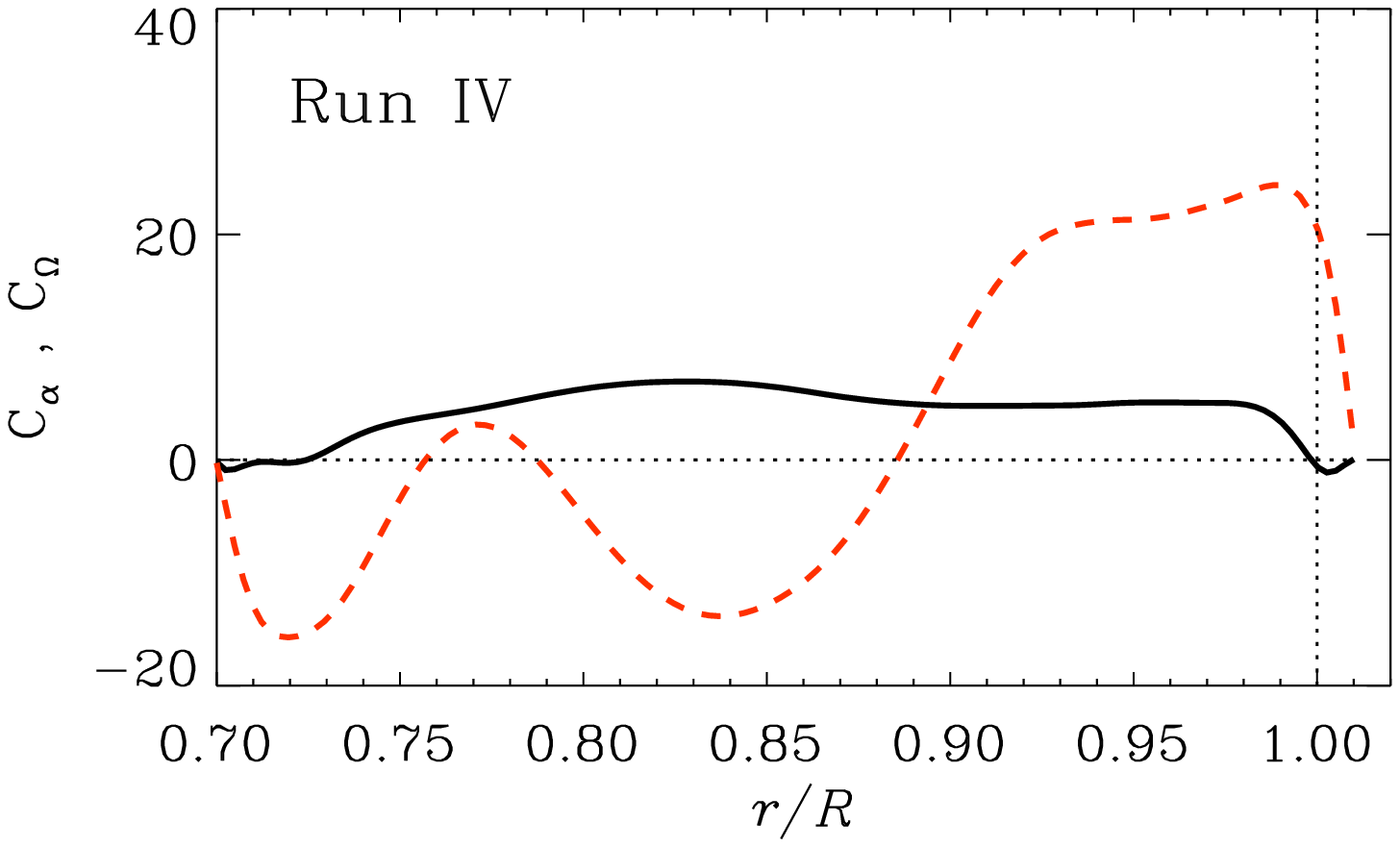}
\end{center}\caption[]{
Local dynamo parameters $C_{\alpha}$ and $C_{\Omega}$ for
Runs~I--IV.
$C_{\alpha}$ (solid black line) and $C_{\Omega}$ (dashed red line)
for 25$^\circ$ latitude in the northern hemisphere as a function of $r$.
}\label{Cparam}
\end{figure*}

\begin{figure}[t!]
\begin{center}
\includegraphics[width=0.45\textwidth]{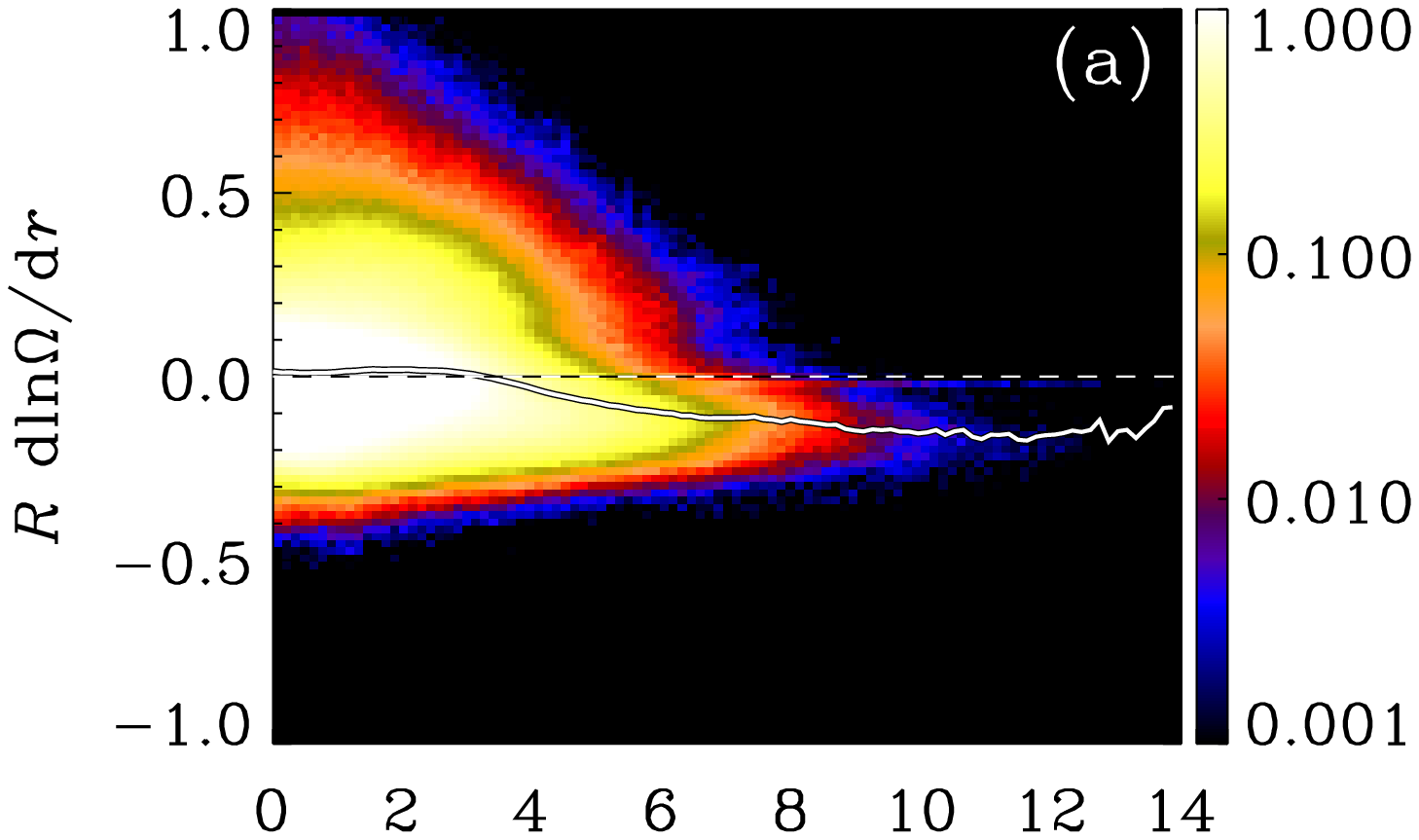}
\includegraphics[width=0.45\textwidth]{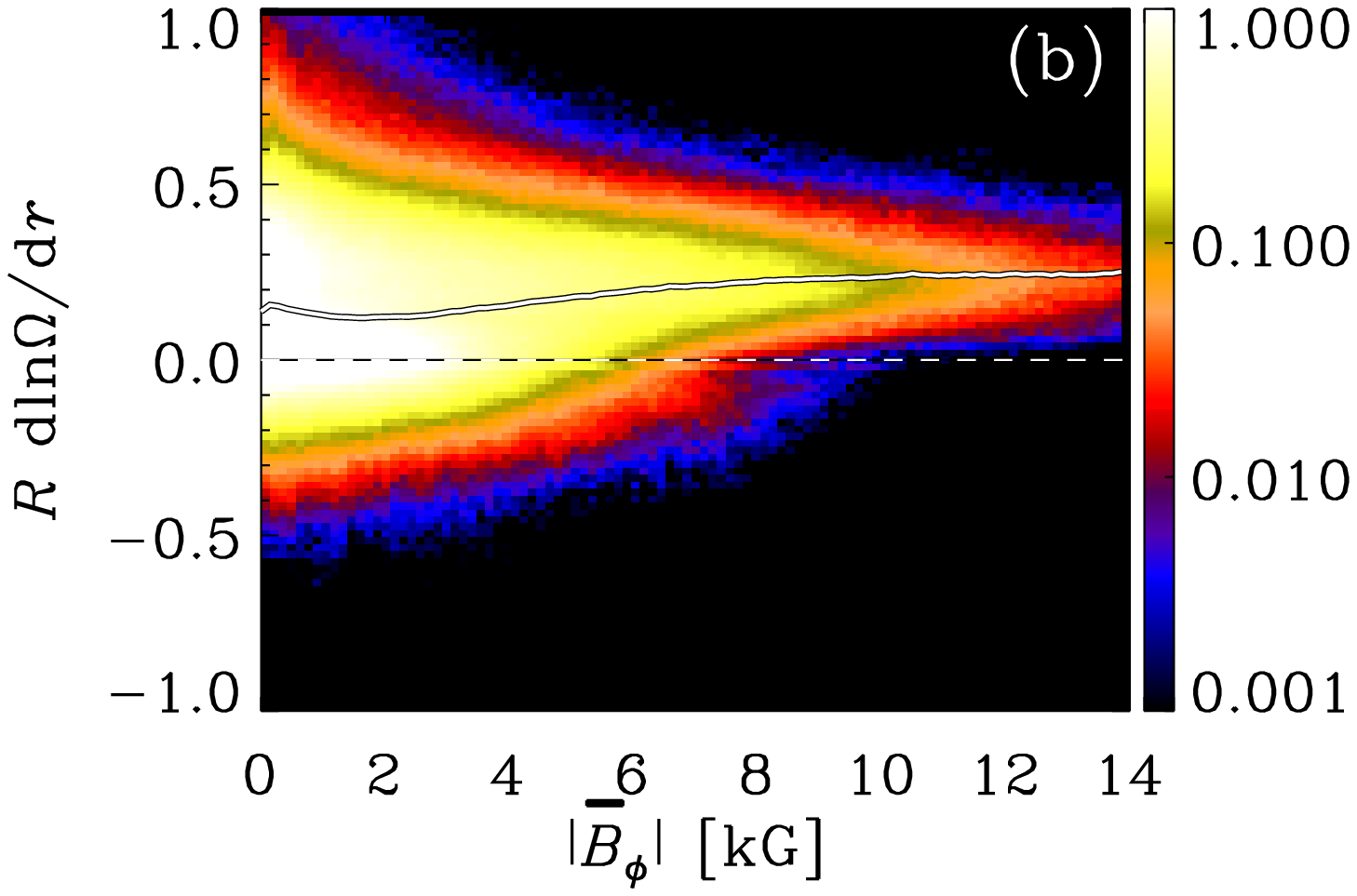}
\includegraphics[width=0.5\textwidth]{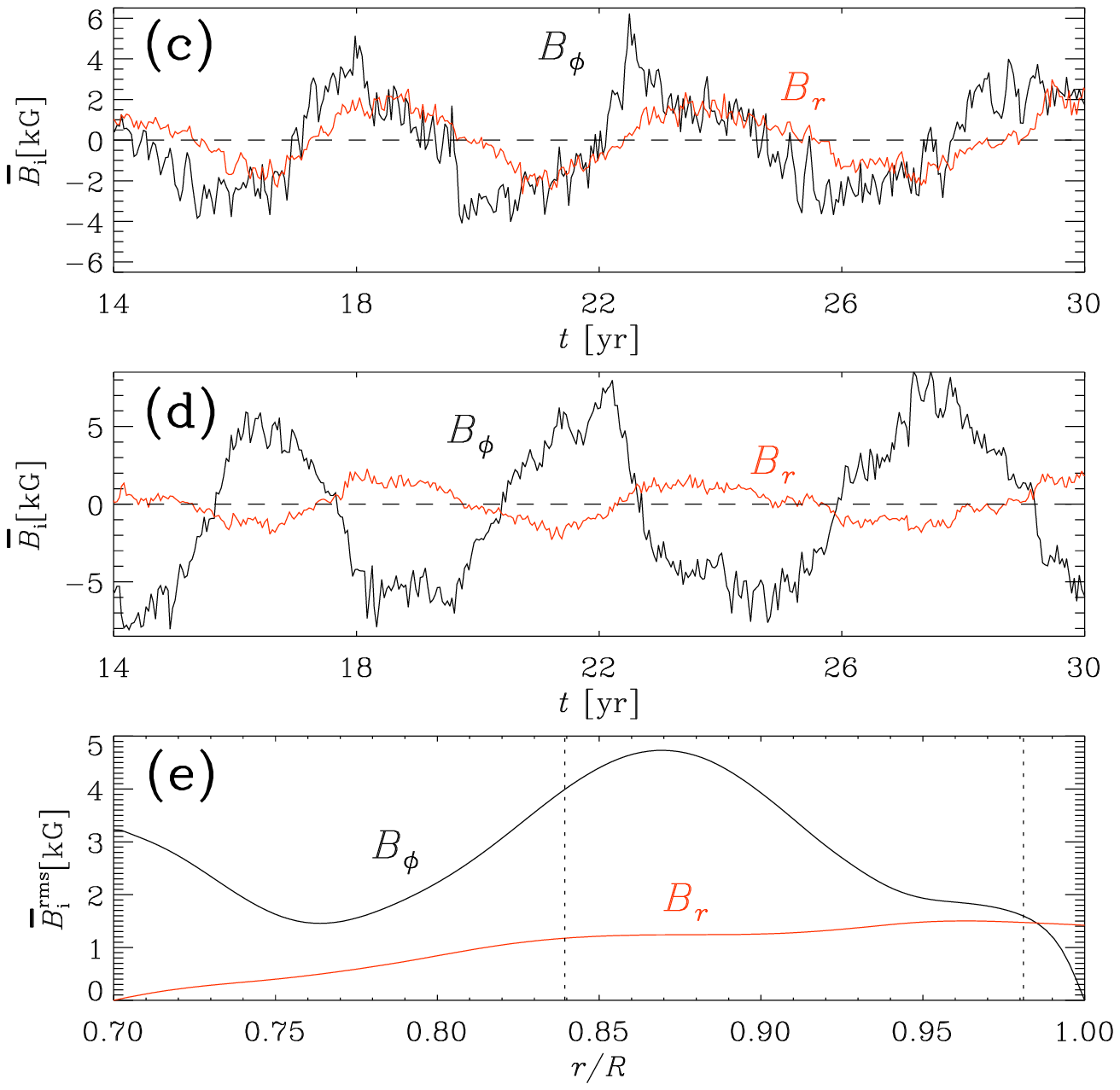}
\end{center}\caption[]{
Panels (a) and (b): correlation of $|\mean{B_{\phi}}|$ from the latitudinal
band $\pm15^{\circ} - \pm40^{\circ}$
and the logarithmic gradient of $\Omega$ for Runs~I (a) and II (b).
Overplotted are the mean (white) and the zero lines
(white-black dashed). 
(c) and (d): phase relation between $\mean{B_\phi}$ (black) and
$\mean{B_r}$ (red) at 25$^{\circ}$ latitude and
$r=0.98\,R$ (c) and at $r=0.84\,R$ (d) for Run~I.
(e): Time-averaged radial dependence of $\Btt$ (black) and
$\Brt$ (red) at 25$^{\circ}$ latitude for Run~I.  }
\label{phases}
\end{figure}

We begin by comparing the evolution of the mean toroidal field $\mean{B_\phi}$
using time--latitude and time--depth diagrams; see \Fig{but}.
In Run~I, $\mean{B_\phi}$ migrates equatorward between $\pm$10$^\circ$ and
$\pm$40$^\circ$ latitude,
and becomes strongly concentrated around $r=0.8-0.9\,R$.
The cycle period is around five years.\footnote{
This agrees with the normalization of \cite{KMB14}.
The difference to the 33\,yr period reported in \cite{KMB12}
is explained by a missing $2\pi$ factor.}
In Run~IV, the evolution of $\mean{B_\phi}$ is similar to Run~I.
Therefore, the blackbody boundary condition is not a necessity for EM.
In both runs, a poleward migrating high-frequency dynamo wave is
superimposed on the EM, as already seen in \cite{KMB12,KMCWB13}.
By contrast, in Run~II, $\mean{B_\phi}$ migrates poleward
between $\pm$10$^\circ$ and $\pm$45$^\circ$ latitude.
The field is strongest at $r=0.85-0.98\,R$ and the
cycle period is about 1.5 yr.
This is clearly shorter than the cycle in
Runs~I and IV, but significantly longer than the superimposed poleward
dynamo wave in those runs.
In Run~III, $\mean{B_\phi}$ has two superimposed field patterns:
one with PM similar in frequency and location to that of Runs~I and IV,
and a quasi-stationary pattern with unchanged field polarity for
roughly 20 yr.
The poleward migrating field appears in the upper 10\%  of the convection
zone, whereas the non-migrating field is dominant in the lower half of the convection zone.

The distribution of $\mean{B_\phi}$ in the meridional plane can
be seen in the top row of \Fig{Diffrot}, where we
plot the rms of the mean toroidal magnetic field, time-averaged over
the saturated stage, $\Btt\equiv\brac{\mean{B_\phi}^2}^{1/2}_t$. 
In Run~I, it reaches 4~kG and is
concentrated at mid-latitudes and mid-depths.
The field structures are aligned with the rotation axis.
Additionally, there is a slightly weaker ($\approx$3~kG) field concentration
closer to the equator and surface.
A similar field pattern can be found in Run~IV, but the field
concentrations are somewhat weaker.
In Run~II, $\Btt$ is concentrated closer to the
surface with a larger latitudinal extent than in Run~I.
The shape of the field structure is predominantly aligned with the
latitudinal direction.
In Run~III, there is some near-surface field enhancement
similar to Run~I, but closer to the equator.
However, the maximum of $\Btt$ is near the
bottom of the convection zone, although at higher latitudes
it occupies nearly the entire convection zone.

Next, we compare the differential rotation profiles of the runs;
see the bottom row of \Fig{Diffrot}.
All runs develop cylindrical contours of constant rotation as a
dominant pattern.
However, Runs~I, III, and IV possess a local minimum of angular velocity, 
implying the existence of a negative $\nablar\Omega$, 
between $\pm$15$^\circ$ and $\pm$40$^\circ$ latitude, which
is the same latitude range where EM was found in Runs~I and IV.
In Run~II, the contours of constant angular velocity are nearly cylindrical,
but with a slight radial inclination, which is more than in Run~I.
This is expected due to the enhanced diffusive heat transport and is also
seen in other global simulations \citep[e.g.][]{BT02,BBBMT08}, where
$\Prat$ is closer to or below unity.
Unlike in Runs~I, III, and IV, there is no local minimum of $\Omega$.
This can be attributed to the higher value of the SGS heat diffusivity in Run~II,
which smoothes out entropy
variations, leading to a smoother rotation profile via the baroclinic term in the thermal wind
balance \citep[see corresponding plots and discussion in][]{WKMB13b}.

Furthermore, we calculate the local dynamo numbers 
\begin{equation}
C_\alpha={\alpha\, \Delta R\over\etatz} ,\quad C_\Omega={\nablar\Omega\, \Delta R^3\over\etatz},
\end{equation}
where $\etatz=\alpha_{\rm MLT} H_p \urms(r,\theta)/3$ is the estimated turbulent
diffusivity with the mixing length parameter $\alpha_{\rm MLT}=5/3$,
the pressure scale height $H_p$, the turbulent
rms velocity $\urms(r,\theta)$, and $\alpha(r,\theta)$ is estimated using
\Eq{alpha}; see also \cite{KMCWB13}.
In \Fig{Cparam} we plot $C_\alpha$ and $C_\Omega$ as functions of
radius for 25$^\circ$ latitude for Runs~I--IV.
The $C_\alpha$ profiles in all the runs are similar: the quantity
is almost always positive, except for a narrow and weak dip to
negative values at the very bottom of the simulation domain.
The only two exceptions are Runs~III and IV, where the cooling 
layer causes $C_\alpha$ to decrease already below (Run~III) or just
above (Run~IV) the surface, becoming even
weakly negative there.
The reason is a sign change of the kinetic helicity caused by the
sign change of entropy gradient.
The $C_\Omega$ profiles are similar for Runs~I, III, and IV. 
There are
two regions of negative values in the lower and middle part
of the convection zone, with positive values near the surface.
In the middle of the convection zone, these profiles
coincide with clearly positive values of $C_\alpha$, as required for EM.
For Run~II, the profiles of $C_\Omega$ are markedly different: despite the
negative dip at the bottom of the convection zone, the values of $C_\Omega$
are generally positive and larger in magnitude than for Runs~I, III, and IV.
This suggests PM throughout most of the convection zone.

To investigate this in more detail, we calculate the migration
direction $\ssss_{\rm mig}$ as \citep{Yos75}
\begin{equation}
\ssss_{\rm mig}(r,\theta)=-\alpha \eee_\phi\times\nab\Omega,
\label{eqm}
\end{equation}
where $\eee_\phi$ is the unit vector in the $\phi$-direction.
Note that this and our estimated $\alpha(r,\theta)$ using \Eq{alpha} is
a strong simplification of the, in general, tensorial properties.
In all of our runs, $\alpha$ is on average positive (negative) in the
northern (southern) hemisphere.

The migration direction $\ssss_{\rm mig}$ is plotted in the top row
of \Fig{Diffrot} for the northern hemispheres of Runs~I--IV.
The white arrows show the calculated normalized migration direction on
top of the color coded $\Btt$ with black contours indicating $\Btt$
= 2.5\, kG.
In Runs~I and IV, \Eq{eqm} predicts EM in the region
where the mean toroidal field is the strongest.
This is exactly how the toroidal field is observed to behave
in the simulation at these latitudes and depths, as seen from \Fig{but}.
The predicted EM in this region is due to $\alpha>0$ and $\nablar\Omega<0$.
Additionally, in a smaller region of strong field closer to the surface
and at lower latitudes, the calculated migration direction is poleward.
This coincides with the high-frequency poleward migrating field shown in
\Fig{but}.
In Run~II, due to the absence of a negative $\nablar\Omega$ (see the
bottom row of \Fig{Diffrot}), $\ssss_{\rm mig}$ points toward the
poles 
in most of the convection zone, in particular in the region where the
field is strongest; see the top row of \Fig{Diffrot}.
Here the calculated migration direction agrees with the actual
migration in the simulation; see \Fig{but}.
In Run~III, there exists a negative $\nablar\Omega$, but in the
region where the toroidal field is strongest, the calculated migration
direction is inconclusive.
There are parts with equatorward, poleward, and even radial migration.
This can be related to the quasi-stationary toroidal field seen in
\Fig{but}.
However, in the smaller field concentration closer to the
surface and at lower latitudes the calculated migration direction is also
poleward, which seems to explain the rapidly poleward migrating
$\mean{B_{\phi}}$ of Run~III (\Fig{but}).
This agreement between calculated and actual migration directions
of toroidal field implies that the EM in the runs of
\cite{KMB12,KMCWB13} and in Runs~I and IV can be ascribed to an
$\alpha\Omega$ dynamo
wave traveling equatorward due to a local minimum of $\Omega$. 

To support our case, we compute a two-dimensional histogram of $|\mean{B_{\phi}}|$
and $\nablar\Omega$
in a band from $\pm15^{\circ}$ to $\pm40^{\circ}$ latitude for
Runs~I and II; see \Fig{phases}(a)--(b).
For Run~I, the strong ($>$5~kG) fields correlate markedly with negative
$\nablar\Omega<0$.
For Run~II, the strong fields are clearly correlated with positive 
$\nablar\Omega<0$.
These correlations have two implications: first, strong fields in
these latitudes are related to and most likely generated by radial
shear rather than an $\alpha$ effect.
Second, the negative shear in Run~I is related to and probably the cause
of the toroidal field migrating equatorward and the positive shear in
Run~II is responsible for PM.

These indications resulting from the comparison of four different
simulation models lead us to conclude that the dominant dynamo mode of
all models is of $\alpha\Omega$ type, and not, as suggested by \cite{KMCWB13},
an oscillatory $\alpha^2$ dynamo.
They based their conclusion on the following three indications.
(1) The two local dynamo numbers, $C_\alpha$ and
$C_\Omega$, had similar values; see Figures 11 and 12 of
\cite{KMCWB13}.
However, due to an error, a one-third factor was missing in the
calculation of $C_\alpha$, so our values are now three times smaller;
see \Fig{Cparam}.
(2) The phase difference of $\approx\pi/2$
between $\mean{B_\phi}$ and $\mean{B_r}$ was observed, which agrees with that of an
$\alpha^2$ dynamo,
as demonstrated in Figure~15 of \cite{KMCWB13}.
As shown in Figures~\ref{phases}(c) and (d), this is only true close to the
surface ($r=0.98\,R$).
At mid-depth ($r=0.84\,R$), where $\Btt$ is strong, the
phase difference is close to $3\pi/4$, as expected for an
$\alpha\Omega$ dynamo with negative shear; see Figure 15(e) of
\cite{KMCWB13}.
(3) Poloidal and toroidal fields had similar strengths, as was shown
in Figures 15(a) and (b) of \cite{KMCWB13}.
Again, however, this is only true near the surface ($r=0.98\,R$), where
$B_\phi$ has to decrease due to the radial field boundary condition.
As shown in \Fig{phases}(e), $\Btt$ and $\Brt$ are
comparable only near $r=0.98\,R$, whereas in the rest of the convection
zone, $\Btt$ is around five times larger than $\Brt$.
It is still possible that there is a subdominant $\alpha^2$ dynamo
operating close the surface causing the phase and strength relation
found in \cite{KMCWB13}.

Comparing our results with \cite{ABMT13}, their differential rotation
profile possesses a similar local minimum of $\Omega$ as our Runs~I,
III, and IV; see their Figure~2(b).
This supports the interpretation that an $\alpha\Omega$ dynamo
wave is the cause of EM also in their case.

Even though the input parameters are similar to those of Runs~I and IV,
in Run~III $\mean{B_\phi}$ does not migrate toward the equator.
The only difference between Runs~III and IV is the higher surface temperature
in the former (\Fig{strat}(a)).
As seen from \Fig{strat}(c),
this leads to a suppression of turbulent velocities
and a sign change of $\alpha$ close to the surface in those latitudes,
where EM occurs in Runs~I and IV; see also \Fig{Cparam}.
One of the reasons might be the fact that the sign changes.
This suppresses the dynamo cycle and causes a quasi-stationary field.
Another reason could be $C_\Omega$ in Run III being stronger at the bottom of
the convection zone than in the middle (in contrast to Runs~I and IV;
see Figure \ref{Cparam}), which implies a preferred toroidal field generation near
the bottom, where the migration direction is not equatorward; see the top
row of \Fig{Diffrot}.

\section{Conclusions}

By comparing four models of convectively driven dynamos,
we have shown that the EM found in the work of \cite{KMB12} and in
Run~IV of this Letter as well as the PM in Runs~II and III can be
explained by the Parker--Yoshimura rule.
Using the estimated $\alpha$ and determined $\Omega$ profiles
to compute the migration direction predicted by this rule, we obtain
qualitative agreement with the actual simulation in the regions where
the toroidal magnetic field is strongest.
This result and the phase difference between the toroidal and poloidal fields
imply that the mean field evolution in these global convective dynamo
simulations can well be described by an $\alpha\Omega$ dynamo with a
propagating dynamo wave.
We found that the radiative blackbody boundary condition is not
necessary for obtaining an equatorward propagating field.
Even though the parameter regime of our simulations might be far
away from the {\it real} Sun, analyzing these simulations, and
comparing them with, e.g., mean-field dynamo models, will lead to a
better understanding of solar and stellar dynamos and their cycles.

\acknowledgments
We thank Matthias Rheinhardt for useful comments on the manuscript.
The simulations have been carried out on supercomputers at
GWDG,
on the Max Planck supercomputer at RZG in Garching and in the facilities
hosted by the CSC---IT Center for Science in Espoo, Finland, which are
financed by the Finnish ministry of education.
This work was partially funded by the Max-Planck/Princeton Center for
Plasma Physics (J.W) and supported in part by the
Swedish Research Council grants Nos\ 621-2011-5076 and 2012-5797
(A.B.),
and the Academy of Finland
Centre of Excellence ReSoLVE 272157 (M.J.K., P.J.K. and J.W.), and
grants 136189 and 140970 (P.J.K).

\bibliography{paper}
\bibliographystyle{apj}

\end{document}